# Influence of La-doping on the magnetic properties of the two-dimensional spin-gapped system $SrCu_2(BO_3)_2$

Lia Šibav[1,2], Tilen Knaflič[1], Graham King[3], Zvonko Jagličić[4,5], Maja Koblar[1], Kirill Yu. Povarov[6], Sergei Zvyagin[6], Denis Arčon[1,7], and Mirela Dragomir[1,2*]

[1] Jožef Stefan Institute, Jamova cesta 39, 1000 Ljubljana, Slovenia
[2] Jožef Stefan International Postgraduate School, Jamova cesta 39, 1000 Ljubljana, Slovenia
[3] Canadian Light Source, 44 Innovation Blvd, Saskatoon, SK S7N 2V3, Canada
[4] Institute of Mathematics, Physics and Mechanics, Jadranska ulica 19, 1000 Ljubljana, Slovenia
[5] Faculty of Civil and Geodetic Engineering, University of Ljubljana, Jamova cesta 2, 1000 Ljubljana, Slovenia
[6] Dresden High Magnetic Field Laboratory and Würzburg-Dresden Cluster of Excellence ct.qmat, Helmholtz-Zentrum Dresden-Rossendorf, 01328 Dresden, Germany
[7] Faculty of Mathematics and Physics, University of Ljubljana, Jadranska ulica 19, 1000 Ljubljana, Slovenia

*Corresponding author: mirela.dragomir@ijs.si

## Abstract

Aliovalent doping of the two-dimensional dimer antiferromagnet $SrCu_2(BO_3)_2$ has long been proposed as a potential route toward realizing resonating valence bond (RVB) superconductivity in this system; however, experimental progress has remained limited. This study explores the effects of La-doping on the ground state of $SrCu_2(BO_3)_2$ and reports the first flux growth of $Sr_{1-x}La_xCu_2(BO_3)_2$ single crystals with nominal doping levels up to $x = 0.15$. Powder X-ray diffraction and energy-dispersive X-ray spectroscopy confirm the successful incorporation of La on the Sr sites within the same tetragonal $I\bar{4}2m$ structure, although the effective doping was limited to approximately 50% of the nominal concentration. La-doping induces systematic changes in the magnetic properties, with a reduction of the effective spin gap, $\Delta$, from 28.2 K to 20.3 K for $x = 0.15$, as determined from the low-temperature magnetic susceptibility. The X-band electron spin resonance measurements reveal the emergence of unpaired $Cu^{2+}$ spins in La-doped $SrCu_2(BO_3)_2$ single crystals, which develop antiferromagnetic correlations below ~5.5 K. These findings corroborate the breaking of the local spin-dimers induced by La-doping. Despite this, no superconductivity is observed across the entire doping range studied.

The present study demonstrates that at low doping levels, electron doping locally destabilizes the spin-singlet ground state in $SrCu_2(BO_3)_2$, while the intrinsic spin dynamics of the dimer lattice remain largely preserved.

1. INTRODUCTION

In the last decades, considerable research has been devoted to low-dimensional quantum magnets, in particular to those exhibiting a spin gap in their excitation spectra. In this context, Mott-Hubbard insulators have been extensively investigated due to their connection to cuprate superconductors. The discovery of high-$T_C$ superconductivity in charge-doped cuprates in 1986[1] indicated the potential of chemical doping as an avenue for accessing novel exotic ground states in low-dimensional and frustrated quantum magnets.[2] This milestone spurred extensive theoretical work, proposing mechanisms for unconventional superconductivity, such as the resonating valence bond (RVB) model on the 2D frustrated spin lattice,[3,4] which predicts that upon doping, the Mott-Hubbard gap closes. This allows previously localized spin-singlet pairs (analogous to Cooper pairs) to propagate freely, leading to the onset of superconductivity.

While there have been several studies on gapped one-dimensional (1D) magnetic systems with spin-singlet ground states, such as spin-Peierls,[5] Haldane chains[6] or spin ladders,[7] there are less examples of two-dimensional (2D) spin-gapped systems.[8,9,10] Thus, systems with 2D spin-singlet ground state and frustrated spin configuration are of great interest. One notable representative of such systems is the frustrated antiferromagnet $SrCu_2(BO_3)_2$.[11,12,13] The unique structure of this material consists of $Cu^{2+}$ ions ($S = ½$), which form a two-dimensional orthogonal dimer lattice that stacks along the *c*-axis. Each $Cu^{2+}$–$Cu^{2+}$ pair constitutes a strongly antiferromagnetically coupled dimer within the *ab*-plane, but the next-nearest exchange to the neighboring dimers is comparable in strength. The resulting network maps directly onto the Shastry–Sutherland model, describing a frustrated quantum spin system characterized by competing intra-dimer ($J$) and inter-dimer ($J'$) antiferromagnetic exchange interactions.[14,15,16] Due to the orthogonal arrangement of adjacent dimers, when the ratio $J'/J$ is sufficiently small, i.e. $J'/J < 0.69$, the model takes an exact ground state comprising a product of singlet states on each dimer.[13] For intermediate values $0.86 > J'/J > 0.69$, a gapped plaquette singlet state, also known as a plaquette valence bond solid, emerges.[17] For $J'/J > 0.86$, the system transitions to a Néel antiferromagnetic state,[17] with a quantum spin liquid phase possibly stabilized between the gapped plaquette-singlet and magnetically ordered regimes.[18]

Experimental values for $J'/J = 0.68$[13] locate $SrCu_2(BO_3)_2$ very close to the quantum phase boundaries either to a Néel ordered state or to the RVB plaquette singlet state. Various experiments provide firm evidences for a dimer spin-singlet ground state with the gap $\Delta = 34(1)$ K.[19,20,21] Magnetization curves collected at temperatures well below the gap, show quantized plateaus at 1/3, 1/4, or 1/8 of the Cu saturation moment due to the localized nature of excited triplets.[12,22] The origin of these plateaus was attributed to the nearly localized nature of the triplet excitations.[13]

Both external pressure and chemical doping have proven effective in tuning the magnetic ground state of $SrCu_2(BO_3)_2$; pressure reduces the spin gap by modifying exchange interactions and pushing the system closer to phase boundaries,[23,24] while magnetic dilution – such as Mg substitution for Cu – suppresses the spin gap by disrupting the singlet dimer order.[25] According to theoretical predictions, a potential RVB plaquette singlet state might be realized as the ground state of $SrCu_2(BO_3)_2$ upon A-site electron or hole doping – specifically, by substituting the interlayer $Sr^{2+}$ with an aliovalent ion, $M^{3+}$ or $M^+$, respectively, resulting in $Sr_{1-x}M_xCu_2(BO_3)_2$.[26,27,28] However, despite its potential, A-site chemical substitution in $SrCu_2(BO_3)_2$ remains remarkably difficult. Previous studies on polycrystalline samples with various dopants, i.e., Ca, Ba, Al, La, Na, or Y, showed only subtle structural and magnetic changes as the effective doping concentrations are consistently low. Moreover, challenges with efficient dopant incorporation and impurity phases highlighted the need for single crystals.[29,30] The optical floating zone method enabled the growth of doped single crystals with dopants such as Ba, Na, or La [31,32,33] but these efforts echoed similar challenges: crystal growth was slow, highly sensitive to chemical composition and growth conditions, and often resulted in multi-grain samples. Each dopant significantly altered the growth parameters, making reproducible, uniform doping difficult, even at low dopant concentrations. Similar issues were reported for B-site doping with isovalent, non-magnetic ions,[25,34,35,36] emphasizing the need for further research and alternative growth methods.

Motivated by reports that La induces the most significant reduction of the spin gap among the A-site dopants,[30] this work focuses on La-doping of $SrCu_2(BO_3)_2$. Previous investigations of both polycrystalline and single-crystalline samples—such as those prepared by Liu *et al.*[30] and Dabkowska *et al.*[32] respectively—were limited in doping range reaching only up to nominal $x$ = 0.10, and hindered by complications including secondary phase formation and instabilities of the crystal growth. In this work, a flux method, previously used for the growth of submillimeter single crystals of undoped $SrCu_2(BO_3)_2$ using $LiBO_2$ as a flux,[11] was optimized to grow plate-like $Sr_{1-x}La_xCu_2(BO_3)_2$ single crystals (with $x$ = 0.02, 0.03, 0.04, 0.05, 0.10, and 0.15), reaching lateral sizes up to 3 mm. La incorporation into the parent structure was unambiguously confirmed by powder X-ray diffraction (PXRD) and energy-dispersive X-ray spectroscopy (EDS). It was found that the effective doping reached ~50% of the nominal value. The limited incorporation of La nonetheless induced progressive changes in magnetic behavior. A gradual reduction of the effective spin gap—from 28.2 K to 20.3 K at $x$ = 0.15—was observed, accompanied by the emergence of dimer-free $Cu^{2+}$ spins at low temperatures. X-band and high-field electron spin resonance (ESR) spectroscopy further showed evidence of antiferromagnetic exchange interactions between these spins below 5.5 K, which primarily accounts for the observed effective spin-gap reduction at low doping levels, while the intrinsic dimer lattice spin dynamics remain largely preserved.

## 2. EXPERIMENTAL
### 2.1.1. Synthesis
**Solid-state synthesis**

Polycrystalline $Sr_{1-x}La_xCu_2(BO_3)_2$ samples with nominal $x$ concentrations of 0, 0.01, 0.02, 0.03, 0.04, 0.05, 0.10 and 0.15, were first prepared using a conventional solid-state method.[25] High-purity $SrCO_3$ (Alfa

Aesar, 99.994%), CuO (Aldrich, 99.99%), $H_3BO_3$ (Alfa Aesar, 99.9995%) and $La_2O_3$ (Alfa Division, 99.99%) were used as starting materials. Prior to use, $La_2O_3$ was pre-annealed in air at 1000 °C for 48 hours. The corresponding reactions can be described by the following equations:

$SrCO_3 + 2\ CuO + 2\ H_3BO_3\ \rightarrow\ SrCu_2(BO_3)_2 + CO_2 + 3\ H_2O$  (1)

$1-x\ SrCO_3 + 2\ CuO + 2\ H_3BO_3 + x/2\ La_2O_3\ \rightarrow\ Sr_{1-x}La_xCu_2(BO_3)_2 + 1-x\ CO_2 + 3\ H_2O$  (2)

### Single-crystal growth

Single-crystal growth experiments were performed by optimizing a previously reported flux method[11] for growing undoped $SrCu_2(BO_3)_2$ for structure solution using lithium metaborate, $LiBO_2$, as the flux. Here, blue $Sr_{1-x}La_xCu_2(BO_3)_2$ powders of nominal doping concentrations $x$ = 0, 0.02, 0.03, 0.04, 0.05, 0.10 and 0.15 that resulted from solid-state reactions were used as starting materials.

In a typical flux growth experiment, between 0.25–0.50 g of polycrystalline $Sr_{1-x}La_xCu_2(BO_3)_2$ was hand-homogenized together with the flux in an agate mortar at material-to-flux mass ratios of 2:1, 3:1, 4:1, 5:1, 7:1 or 10:1. The homogenized mixture was transferred into a 20 or 30 mL platinum crucibles, covered with a platinum lid, and positioned inside an alumina crucible covered with an alumina cap. The crucible was then heated to 875 °C in a muffle furnace at a heating rate of 100 °C/h and left to dwell for 1, 2 or 3 hours, followed by slow cooling to 600 °C with cooling rates of 1, 2, 3, 5 or 10 °C/h before being air quenched to room temperature. A schematic representation of the experimental setup and the growth process is shown below (**Figure 1**). The resulting blue thin plate-like single crystals with lateral sizes up to 3 mm were separated from the crucible using deionized water in an ultrasonic bath.

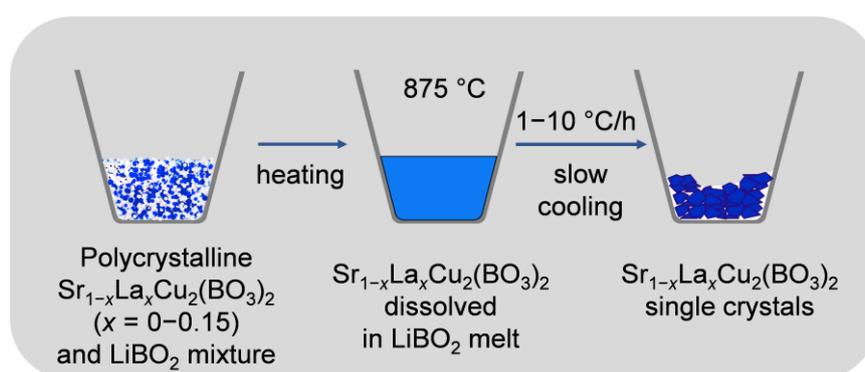

**Figure 1** A schematic representation of the experimental setup and process for the $Sr_{1-x}La_xCu_2(BO_3)_2$ flux growth, illustrating three key steps: (left) preparation of the initial polycrystalline mixture of $Sr_{1-x}La_xCu_2(BO_3)_2$ and $LiBO_2$ flux, (middle) dissolution of the mixture in molten flux at a dwell temperature of 875 °C, and (right) formation of $Sr_{1-x}La_xCu_2(BO_3)_2$ single crystals during slow cooling.

### 2.1.2. Characterisation

#### Powder X-ray diffraction (PXRD)

The phase composition of undoped and La-doped polycrystalline samples obtained after the solid-state synthesis, as well as powders from crushed single crystals, was first probed using laboratory powder X-ray diffraction (PXRD) with a Panalytical X'Pert Pro powder diffractometer and Cu-Kα1 radiation in the 20–120° 2θ range with a step size of 0.016° and a counting time of 300 s per step. Crushed single crystals were also analyzed by synchrotron PXRD; the data were collected at the Brockhouse high energy wiggler beamline[37] at the Canadian Light Source (CLS) using an area detector and λ = 0.3502 Å radiation with Ni as a calibrant. Single crystals of each nominal doping concentration were crushed, placed into Kapton capillaries, measured in 2θ range 1–26° and binned into 0.01° steps.

A Rietveld least-square method was used for structural refinements, which were performed using the program GSAS-II.[38]

#### Scanning Electron Microscopy (SEM) and Energy-dispersive X-ray spectroscopy (EDS)

For chemical analysis and microstructural investigation, $Sr_{1-x}La_xCu_2(BO_3)_2$ single crystals were mounted on a carbon tape and carbon coated using a sputter coater Balzers SCD 050. The SEM imaging and EDS compositional analyses of the crystals were performed on two instruments: a Thermo Fisher Quanta 650 ESEM equipped with an energy-dispersive X-ray spectrometer (Oxford Instruments, AZtec Live, Ultim Max SDD 65 mm$^2$) and a field-emission-gun scanning electron microscope (FE-SEM; JEOL JSM-7600) equipped with an energy-dispersive X-ray spectrometer (EDS; INCA Oxford 350 EDS SDD) and electron backscatter diffraction (EBSD) The accelerating voltage used was 20 kV in all cases.

#### Magnetic susceptibility

Magnetic susceptibility measurements were performed on a Quantum Design MPMS3 (undoped $SrCu_2(BO_3)_2$) and MPMS-XL-5 (La-doped $SrCu_2(BO_3)_2$) SQUID magnetometers in a temperature interval 2–300 K in a static magnetic field of 1 kG. A selected undoped $SrCu_2(BO_3)_2$ single crystal was fixed to a quartz holder with Apiezon-N grease and inserted into the magnetometer. Several La-doped $SrCu_2(BO_3)_2$ single crystals were measured as polycrystalline samples at the bottom of a Wilmad 4 mm Suprasil® ESR sample tube. The temperatures corresponding to the susceptibility maxima, $T_{max}$, were determined by fitting the susceptibility data near the peak to a simple Lorentzian function.

#### Electron spin resonance spectroscopy (ESR)

Continuous wave (CW) electron spin resonance spectroscopy experiments were performed using a conventional Bruker E500 spectrometer, operating in the X-band at the resonant frequency $v_L$ of 9.37

GHz. The spectrometer was equipped with a Varian TEM104 dual cavity resonator, an Oxford Instruments ESR900 cryostat and an Oxford Instruments ITC503 temperature controller.

The measurements were conducted from room temperature down to 4 K with a typical modulation amplitude of 5 G, a modulation frequency between 50 and 100 kHz, and microwave power of 1–2 mW. For each measurement, approximately 30 mg of polycrystalline undoped $SrCu_2(BO_3)_2$ and 5–30 mg of La-doped small single crystals (measured as quasi-polycrystalline samples) were placed in 4 mm Suprasil® quartz ESR tubes (Wilmad). For samples with $x$ = 0.10 and $x$ = 0.15, the ESR spectrometer was equipped with a Bruker 4122SHQE cylindrical resonator operating at 9.4 GHz. Due to a higher sensitivity of this resonator, the latter two samples were measured using a low microwave power of 0.1 mW.

High-field ESR measurements were performed at the High Magnetic Field Laboratory (HLD), Helmholtz-Zentrum Dresden Rossendorf (HZDR) employing a transmission-type ESR spectrometer (similar to that described in ref.[39]) in magnetic fields up to 16 T. Multiple crystals with nominal La concentration of 0.03 were stacked using vacuum grease. A set of VDI microwave sources was used, allowing to probe magnetic excitations at different frequencies in the range from 140 to 495 GHz. Prior to the measurements of ESR spectra, the sample was slowly cooled to the base temperature $T$ = 1.7−2.9 K.

## 3. RESULTS AND DISCUSSION

### 3.1.1. Synthesis and crystal growth

Solid-state synthesis of $Sr_{1-x}La_xCu_2(BO_3)_2$ with $x$ = 0.01, 0.02, 0.03, 0.04, 0.05, 0.10, and 0.15 predominantly results in the tetragonal $SrCu_2(BO_3)_2$ phase with the $I\bar{4}2m$ space group, as confirmed by laboratory PXRD (**Figure 2**). A small amount of CuO is also present, consistent with previous reports on undoped polycrystalline samples.[11,40] A secondary La phase, $LaBO_3$, becomes noticeable when $x \geq 0.01$, suggesting that La does not incorporate into the parent structure but instead forms a La-rich impurity phase. A systematic increase in the concentrations of both $LaBO_3$ and CuO with increasing La content is also observed, supporting this conclusion. This is additionally substantiated by the absence of notable peak shifts in the PXRD patterns relative to the undoped sample.

Since a direct solid-state synthesis proved ineffective for incorporating La, single-crystal growth was pursued as an alternative. Given the reported challenges associated with the optical floating zone growth,[32] the flux method emerged as a more practical and accessible approach for growing these crystals. This technique does not require highly specialized equipment and was considered likely to succeed as it had previously been used to grow crystals of the undoped phase.[11,41,42] In a previous attempt to synthesize $SrCu_2(BO_3)_2$[11] crystals intended for structure determination via single-crystal X-ray diffraction (SCXRD), $LiBO_2$ was used as flux. Two other subsequent studies reported the use of another borate flux, namely $Na_2B_4O_7$.[41,42] In the present study, both fluxes were tested in initial experiments; however, only $LiBO_2$ yielded crystals in our setup and was thus chosen for optimization and subsequent growth experiments. The starting materials for these crystal growth experiments were $Sr_{1-x}La_xCu_2(BO_3)_2$ powders prepared via solid-state synthesis.

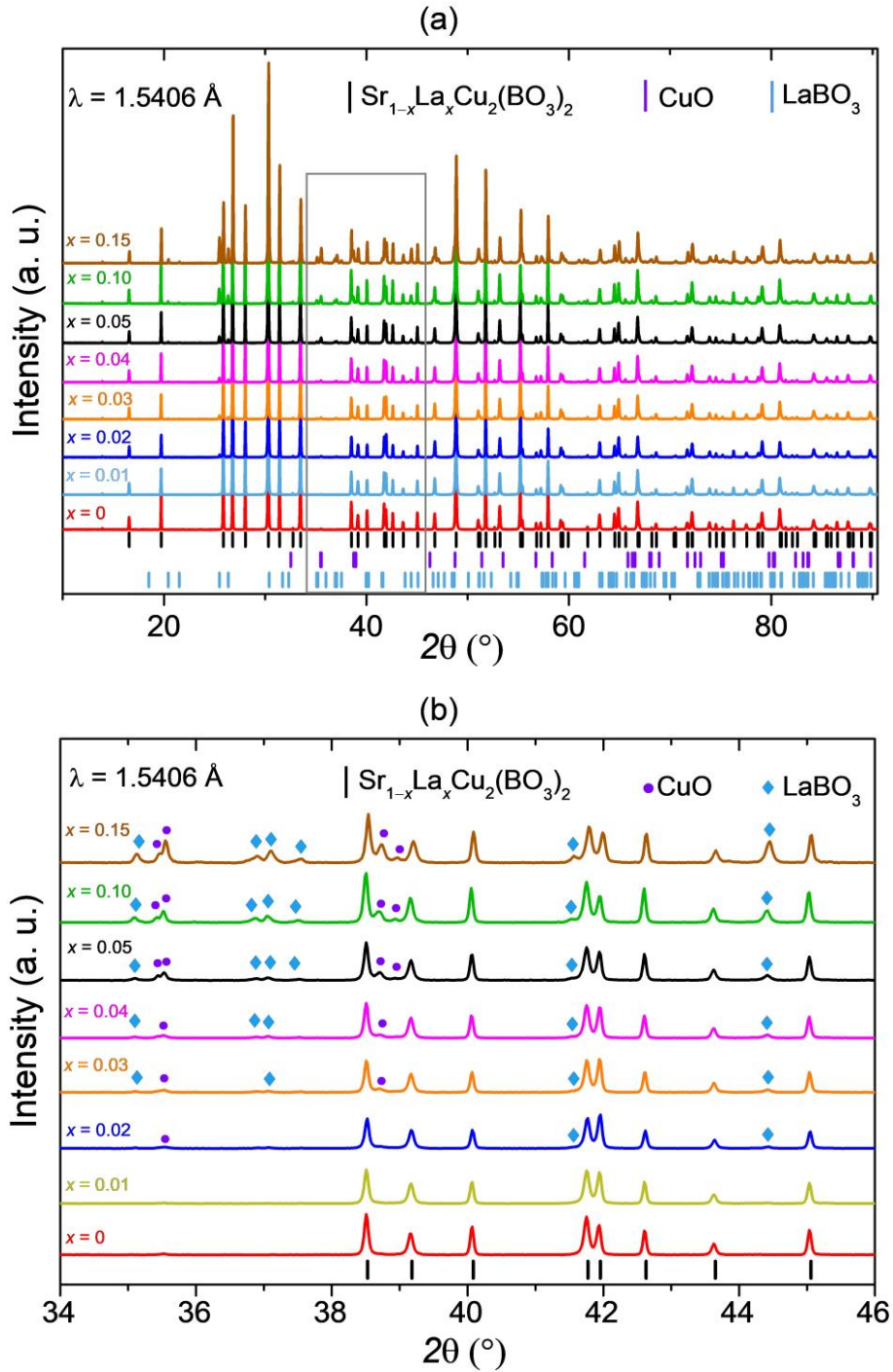

**Figure 2** (**a**) Laboratory PXRD patterns of polycrystalline $Sr_{1-x}La_xCu_2(BO_3)_2$ with $x$ = 0, 0.01, 0.02, 0.03, 0.04, 0.05, 0.10, and 0.15 as synthesized from the thermal solid-state chemistry route. All doped samples exhibit the main $SrCu_2(BO_3)_2$ phase, accompanied by minor amounts of unreacted CuO and $LaBO_3$ as secondary phases. (**b**) Enlarged view of the 34°–46° 2θ range marked with a grey rectangle in a), highlighting the reflections corresponding to the CuO and $LaBO_3$ impurities for improved clarity.

The material-to-flux mass ratio, dwell time, dwell temperature and the cooling rate were optimized to result in blue plate-like undoped and La-doped $SrCu_2(BO_3)_2$ single crystals with lateral sizes of up to 3 mm (**Figure 3**). The optimal conditions within the tested parameters (**Section 2.2**), which yielded the largest crystals, are a material-to-flux mass ratio of 3:1 or 4:1, a dwell time of 2 hours at 875 °C, and cooling from 875 to 600 °C at a rate of 2 °C/h, followed by air quench to room temperature. Material-to-flux mass ratios of 5:1 or higher, as well as 2:1 or lower, result in crystals with minimal sizes. A faster cooling rate of 10 °C/h does not yield single crystals, whereas slower cooling rates of 5 or 3 °C/h produced crystals, albeit smaller compared to those formed at the optimal 2 °C/h rate. Further reduction of the cooling rate to 1 °C/h does not offer any additional benefit in crystal size. In contrast to the dwell time, which shows no notable effect, the material-to-flux mass ratio and cooling rate are identified as key factors influencing the crystal growth. We note that for undoped $SrCu_2(BO_3)_2$, the resulting crystals are significantly larger than those reported in reference [11], corroborating a successful optimization of growth conditions in this study.

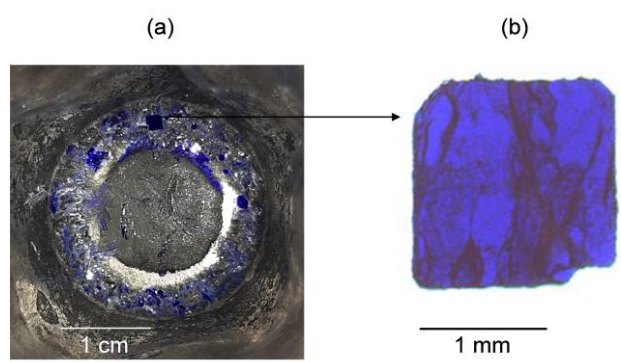

**Figure 3** (**a**) A representative image of as-grown $Sr_{1-x}La_xCu_2(BO_3)_2$ single crystals on the walls of a Pt crucible immediately after air quenching. (**b**) Optical image of a selected single crystal after its separation from the flux.

The incorporation of La in the parent $SrCu_2(BO_3)_2$ structure during single-crystal growth is assessed by PXRD performed on crushed single crystals of $Sr_{1-x}La_xCu_2(BO_3)_2$ crystals with nominal $x$ = 0–0.15 (**Figure 4a**). An absence of La-impurities in normalized synchrotron PXRD patterns indicates a successful incorporation of lanthanum in the parent $SrCu_2(BO_3)_2$ structure upon single-crystal growth. At the highest nominal concentrations, $x$ = 0.10 and 0.15, $LaBO_3$ reappears as a minor impurity, as previously detected in polycrystalline samples prepared by thermal solid-state-chemistry route. Its presence is likely due to residual unincorporated $LaBO_3$ particles remaining on the crystal surfaces, as ultrasonication—used to separate the crystals from the flux after growth—was kept brief to avoid damaging the thin, plate-like crystals. This interpretation is further supported by EDS analysis (discussed in the next section), where $LaBO_3$ particles are observed on the surface of crystals with these nominal doping concentrations. Unreacted CuO is detected in all samples, originating from the crystal surfaces, as suggested by EDS analysis and prior reports.[11]

A detailed summary of Rietveld refinement profiles and resulting structural parameters is available in the Supplementary Information (**Figure S1** and **Table S1, S2**). Rietveld refinement analyses with the $I\bar{4}2m$ space group (**Supplementary Information, Table S1**) reveal a generally decreasing trend in the unit cell parameters and volume as a function of nominal doping concentration (**Figure 4b**). This trend is consistent with expectations based on the ionic radii of the substituting ions, where $Sr^{2+}$ (1.26 Å, coordination number CN = 8) is partially replaced by $La^{3+}$ (1.16 Å, CN = 8). Specifically, the unit cell volume decreases from 537.91(2) Å$^3$ for the undoped sample to 536.68(1) Å$^3$ for the sample with nominal $x$ = 0.15, corresponding to a slight volume reduction of 0.17%, which suggests a minimal but clearly detectable structural distortion. However, an anomaly is observed at $x$ = 0.05, where the unit cell volume shows a small increase compared to the value at $x$ = 0.04. This deviation does not align with the trends observed in the EDS and magnetic susceptibility, which otherwise indicated a higher degree of La incorporation at $x$ = 0.05 compared to $x$ = 0.04. It also contrasts with literature reports on doped crystals with nominal $x$ = 0.05, grown by the optical floating zone method,[32] which consistently show a continuous decrease in unit cell volume relative to both undoped ceramics and single crystals. The observed discrepancy might indicate a possible error in the doping concentration of +/− 0.01.

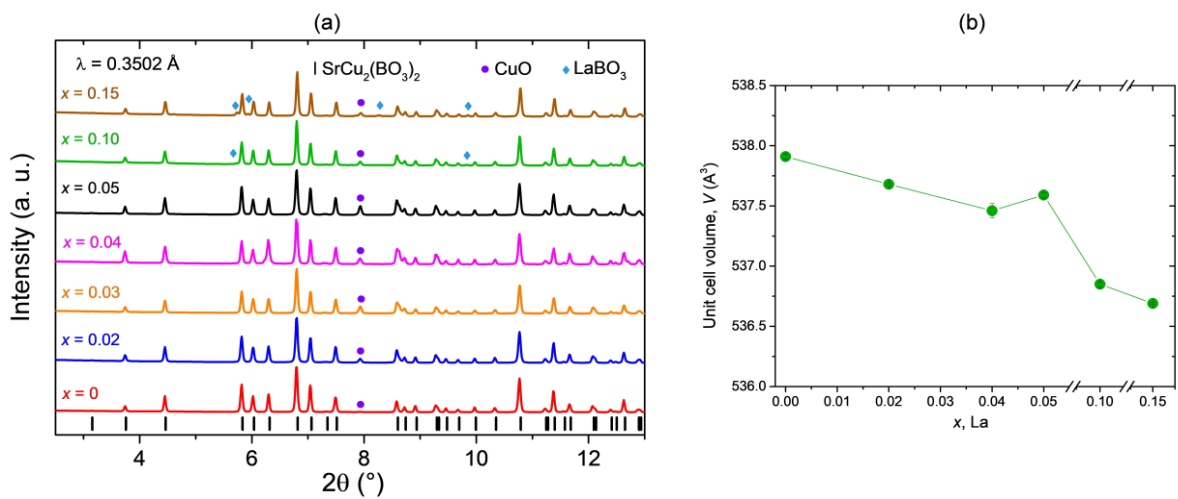

**Figure 4** (a) Synchrotron (λ = 0.3502 Å) PXRD of $Sr_{1-x}La_xCu_2(BO_3)_2$ crushed single crystals with nominal $x$ = 0–0.15. The presence of the main phase, $SrCu_2(BO_3)_2$, is observed along with a small fraction of unreacted CuO. A few peaks corresponding to the $LaBO_3$ impurity are also noticed at the highest two nominal concentrations, $x$ = 0.10 and x = 0.15, with the former being barely detectable. (**b**) The unit cell volume decreases with increasing nominal La-doping, consistent with the smaller ionic radius of $La^{3+}$ compared to $Sr^{2+}$.

### 3.1.2. Scanning Electron Microscopy (SEM) and Energy-Dispersive X-ray spectroscopy (EDS)

Previous results on polycrystalline samples indicated that La does not incorporate in the structure during the solid-state synthesis.[43] This is confirmed also by our findings—see Section 3.1.1. However, SEM-EDS analyses performed on single crystals unambiguously prove the successful incorporation of La into the parent structure of $SrCu_2(BO_3)_2$ during single-crystal growth as lanthanum is consistently detected for all nominal doping concentrations (**Figure 5**).

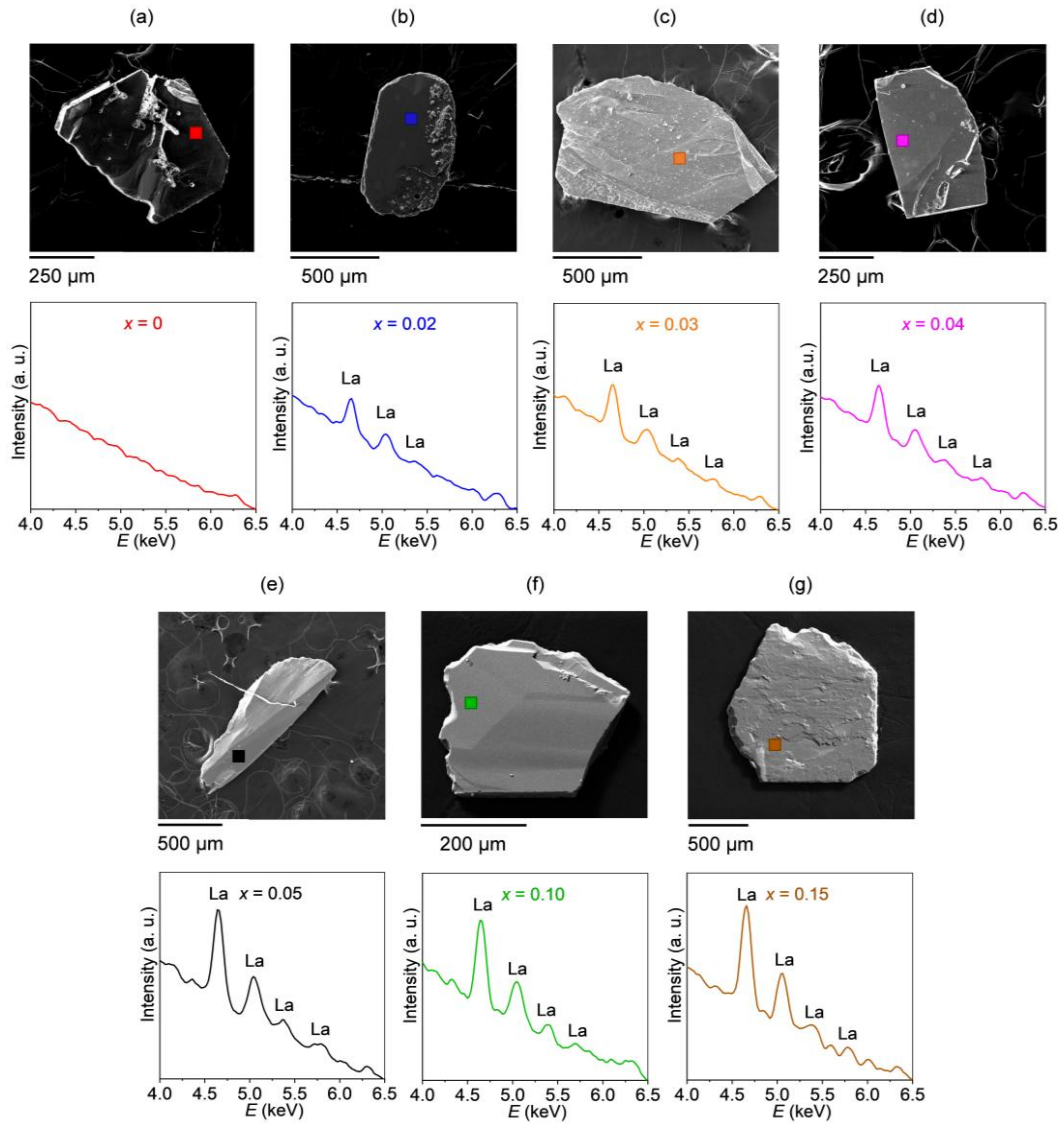

**Figure 5** (**a**–**g**) Representative EDS point analysis spectra for $Sr_{1-x}La_xCu_2(BO_3)_2$ single crystals with $x$ = 0–0.15. The corresponding SEM images for each doping concentration are displayed above the EDS spectra, with squares indicating the selected regions for the point analysis.

The characteristic La emission peaks are clearly observed at 4.65 keV (Lα) and 5.04 keV (Lβ$_1$) in all doped samples. Although two other La peaks, the Lβ$_2$ emission at 5.38 keV and the La emission at 5.89 keV, are less pronounced, they are still detectable for nominal $x \geq 0.03$. Additionally, a systematic increase in La peak intensity is observed with $x$, peaking at $x$ = 0.05 and stagnating for $x$ = 0.10 and 0.15. A similar trend is also observed in elemental mapping analyses, which further reveal a relatively homogeneous distribution of La in all doped crystals. A representative SEM image, elemental maps, and the summed EDS spectrum for a doped crystal with nominal $x$ = 0.05 are shown below (**Figure 6**), with correspondent data for nominal $x$ = 0.04 and 0.10 available in **Supplementary Information, Figure S2**. Impurity particles observed in PXRD – $LaBO_3$ and CuO – are observed on the surfaces of crystals starting from the nominal concentration of $x$ = 0.04 (**Supplementary Information, Figure S2a**). These impurities become most prominent at the highest nominal doping concentration of $x$ = 0.15 (**Supplementary Information, Figure S3**), in agreement with the PXRD results.

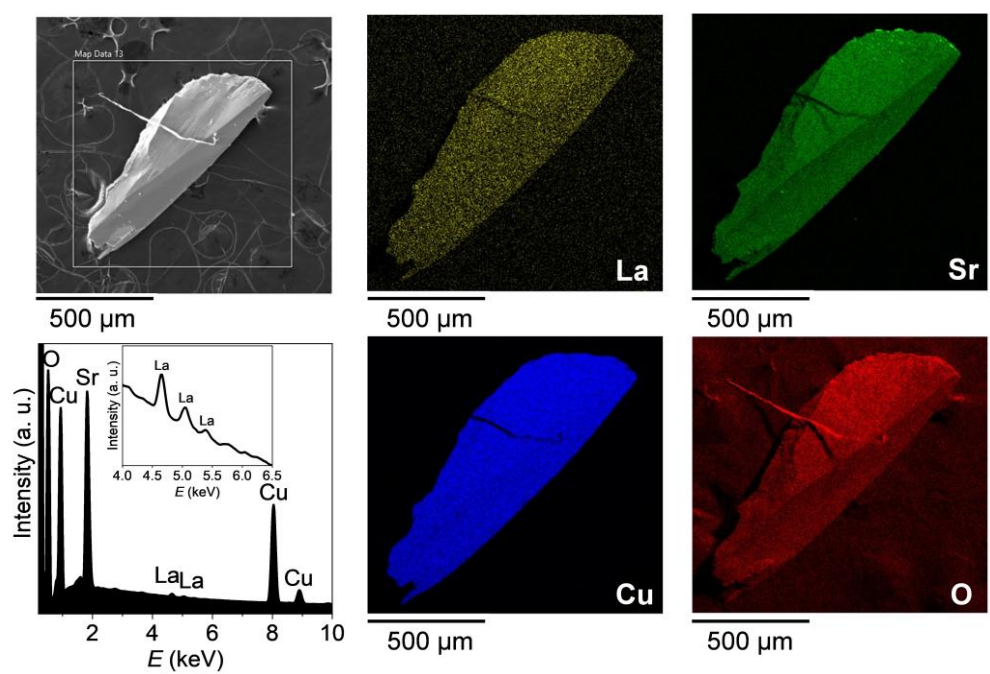

**Figure 6** An SEM image and EDS elemental mapping analysis of a $Sr_{1-x}La_xCu_2(BO_3)_2$ single crystal with nominal $x$ = 0.05 reveal a relatively homogeneous distribution of La within the $SrCu_2(BO_3)_2$ matrix. The corresponding summed EDS spectrum clearly shows the presence of La emission lines. The inset shows a magnified region where the La emission lines are clearly seen.

The average semi-quantitative atomic percentages of Sr, Cu, O, and La in both undoped and La-doped $SrCu_2(BO_3)_2$ single crystals, obtained from SEM-EDS point analyses, are summarized in **Supplementary Information, Table S3**. These values are compared with nominal values, i. e. the intended or theoretical amount of dopant defined by the chemical formulae. The table also includes additional ratios such as La/Cu, La/Sr, and Cu/Sr as well as semi-quantitative estimates of effective La-doping for each nominal concentration. The average La atomic percentage increases with nominal doping, peaking at $x$ = 0.05. However, at $x$ = 0.04, a slight decrease is observed compared to $x$ = 0.03, likely due to small variations in nominal doping. On average, the effective doping — i.e., the actual amount of dopant incorporated into the crystal structure — reaches approximately 50% of the nominal value for $x ≤ 0.05$ (**Supplementary Information, Table S3**). At higher doping levels ($x$ = 0.10 and 0.15), the effective doping remains comparable to that at $x$ = 0.05, indicating a saturation point as the nominal concentrations are further increased. Specifically, the effective doping was estimated to be 2.4(3) mol% for $x$ = 0.10, and 2.5(7) mol% for $x$ = 0.15 compared to 2.6(5) mol% for $x$ = 0.05 (**Supplementary Information**, **Table S3**). Although the effective doping levels for nominal concentrations above $x$ = 0.05 are not significantly surpassed, this may not be reliably captured due to the semi-quantitative nature of EDS and its associated error margins. This limitation could explain the reductions in unit cell volume or spin gap (*vide infra*, Section 3.3), which are still observed at nominal $x$ = 0.10 and 0.15. While these EDS results confirm La incorporation into the $SrCu_2(BO_3)_2$ structure, they also indicate that doping is only partial, saturating slightly above the nominal $x$ = 0.05 at around 2.5 mol%. This finding aligns well with previous studies, [25,32,33] which emphasize the challenges of chemical substitutions in this system.

### 3.1.3. Magnetic susceptibility

Following the successful incorporation of La into the $SrCu_2(BO_3)_2$ structure, we next measured dc magnetic susceptibility in the 2–300 K temperature region at a magnetic field of 1 kG. The behavior of the susceptibility curves (**Figure 7**) is consistent across all the nominal doping concentrations (measured up to $x$ = 0.15) and aligns well with previous reports in pristine and doped $SrCu_2(BO_3)_2$ samples.[30,32,33] Following the high-temperature Curie-Weiss region, a broad susceptibility maximum at temperature $T_{max}$, appears between 15.5 and 18.7 K. The values of $T_{max}$ (**Table 1**) do not show significant variations with $x$, exhibiting only a minor decrease with the increasing nominal doping. At lower temperatures, the susceptibility exhibits a sharp suppression, indicative of the spin gap, followed by a low-temperature Curie–like upturn — a hallmark of unpaired $Cu^{2+}$ $S$ = ½ spins[30,32,33] — but which also includes a thermally activated contribution and a temperature-independent contribution from the ion cores (described by **Equation 3**). The magnitude of this upturn systematically increases with the nominal La-doping concentration and is most pronounced at $x$ = 0.15 (inset, **Figure 7**). The observed magnetic response is in good agreement with previous studies on La-doped single crystals.[32,33] The current study also shows that increasing nominal doping leads to a higher concentration of free $Cu^{2+}$ spins in the system, consistent with a doping-induced $Cu^{2+}$ spin-dimer breaking. A dimer breaking in Sr-site substituted samples was previously observed and directly probed by µSR in ref. [33].

To estimate the evolution of the spin gap with La-doping, low-temperature magnetic susceptibility data are fitted in the interval 2–6 K according to **Equation 3**:[30]

$$\chi = \frac{C'}{T - \theta'} + ae^{\frac{-\Delta}{T}} + \chi_0. \qquad (3)$$

The first term represents intrinsic magnetic impurities, characterized by the Curie constant C' and their weak coupling captured by the Curie-Weiss temperature, θ'. The effective spin gap, Δ, is included in a thermally activated contribution, while $\chi_0$ represents a temperature-independent contribution from ion cores. Due to a high correlation of parameters C' and θ', the fits are performed with θ' fixed at five different values from 0 to −1 (detailed results are available in **Supplementary Information**, **Table S4**), with the fits results for θ' = −0.75 K being chosen as the most representative. The results reveal a gradual closing of the spin gap from 28.2(3) to 20.3(4) K for nominal doping up to *x* = 0.15 (**Table 1**) relative to undoped $SrCu_2(BO_3)_2$. This finding is consistent with the results published in ref. [30], where a similar gap suppression for polycrystalline $Sr_{0.9}La_{0.1}Cu_2(BO_3)_2$ from 21.5 to 14.1 K at 1 T was found. Specifically, the 7.4 K reduction in Δ observed in ref.[30] equals the reduction observed in the current study for the *x* = 0.10 sample.

The concentration of overall dimer-free $Cu^{2+}$ spins in the studied samples is determined by fitting the inverse magnetic susceptibility data to a Curie–Weiss law in the temperature range 1.8–3.5 K (**Supplementary Information, Figure S4**) to obtain the Curie constant, C'. The results (**Table 1**) show a general increase of the Curie constant C' with nominal doping. Specifically, a nominal doping of *x* = 0.10 approximately increases the concentration of dimer-free spins by a factor of approximately 7 compared to the undoped crystal, rising from 0.28(2) to 1.84(5)%, and reaching 2.45(11)% for nominal *x* = 0.15. Crystals of nominal *x* = 0.04 yield 1.36(5)%, which is about 2.5 times higher than the fraction of impurity spins in optical-floating-zone-grown crystals by Dabkowska *et al.*[32] as reported in ref. [33]. By contrast, the increase of impurity spins observed in this study is relatively small compared to Mg-doping.[25]

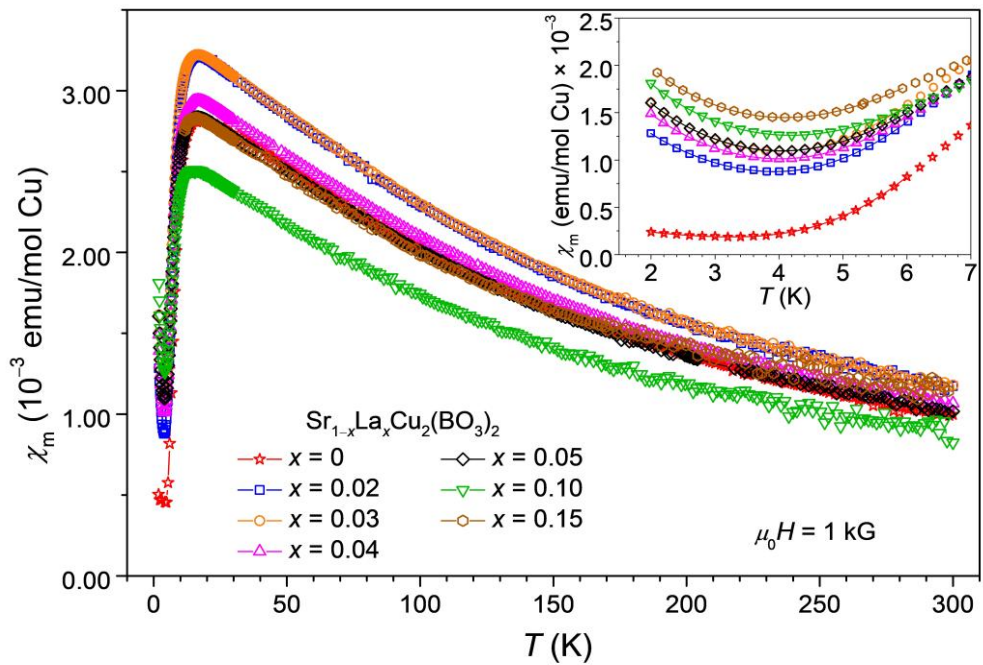

**Figure 7** Temperature dependencies of magnetic susceptibility of $Sr_{1-x}La_xCu_2(BO_3)_2$ single crystals with nominal *x* = 0–0.15, measured at $\mu_0H$ = 1 kG. The inset magnifies the low-temperature region, which exhibits the Curie-Weiss upturn, and includes low-temperature fits based on Equation 3, performed in the 2–6 K range with θ' fixed at −0.75 K.

Table 1 Summary of magnetic parameters for $Sr_{1-x}La_xCu_2(BO_3)_2$ single crystals with nominal $x$ = 0–0.15. The characteristic temperature $T_{max}$ is extracted from Lorentz fits to the magnetic susceptibility in the 15–21 K range. The spin gap is obtained from fits to the low-temperature susceptibility data (2–6 K) using Equation 3. The Curie constant $C'$ is derived from Curie-Weiss fits to inverse susceptibility data at low-temperature (2–3.4 K, **Supplementary Information, Figure S4**). The estimated fraction of free $Cu^{2+}$ spins is derived from these fits.

| $SrCu_2(BO_3)_2$ doping | $T_{max}$ (K) | Spin gap, Δ (K) | C' (emu K / mol Cu, low-$T$ fit 2–3.4 K) | Estimated fraction of free $Cu^{2+}$ spins (%) |
|---|---|---|---|---|
| $x$ = 0 | 18.4(3) | 28.2(3) | $1.05(7) \cdot 10^{-3}$ | 0.28(2) |
| $x$ = 0.02 | 17.05(6) | 24.7(2) | $4.3(2) \cdot 10^{-3}$ | 1.15(6) |
| $x$ = 0.03 | 16.8(2) | 23.2(2) | $5.4(2) \cdot 10^{-3}$ | 1.44(5) |
| $x$ = 0.04 | 16.68(5) | 23.1(3) | $5.1(2) \cdot 10^{-3}$ | 1.36(5) |
| $x$ = 0.05 | 16.45(6) | 22.1(2) | $5.6(2) \cdot 10^{-3}$ | 1.49(6) |
| $x$ = 0.10 | 15.8(2) | 20.8(1) | $6.9(2) \cdot 10^{-3}$ | 1.84(5) |
| $x$ = 0.15 | 15.8(3) | 20.3(4) | $9.2(4) \cdot 10^{-3}$ | 2.45(11) |

### 3.1.4. X-band Electron Spin Resonance Measurements (ESR)

To gain deeper insights into the magnetism of La-doped $SrCu_2(BO_3)_2$, continuous wave X-band electron spin resonance (ESR) spectroscopy, a local probe technique well-suited for detecting paramagnetic states in low-dimensional magnets,[44] is employed next. This study extends the use of ESR spectroscopy to La-doped $SrCu_2(BO_3)_2$ — a system that has received little attention so far. Previous ESR studies have primarily focused on the undoped system,[43,44,45,46,47] with only one report on Mg-doped $SrCu_2(BO_3)_2$.[25] In this study, temperature-dependent ESR spectra for La-doped single crystals with nominal $x$ = 0.02–0.15 were measured down to approximately 4 K, and compared with the ESR data obtained on polycrystalline undoped $SrCu_2(BO_3)_2$ reported in the literature.[25] At high temperatures, the ESR spectra across all nominal doping concentrations closely follow the behavior observed in the undoped system (**Supplementary Information, Figure S5**). The main signal, originating from the dimer lattice, broadens with the decreasing temperature, reflecting the development of spin correlations within the Shastry-Sutherland dimer lattice. At low temperatures, an additional component, only weakly observed in the undoped system at $T \leq 14$ K, emerges in the spectra of doped samples, at $T \leq 35$ K for nominal $x$ = 0.02 and 0.03, and $T \leq 50$ K for higher nominal concentrations. As the nominal doping increases, this component becomes progressively more pronounced (**Supplementary Information, Figure S6**). Its ESR signal intensity scaling with $x$ aligns well with the increase of the Curie-Weiss upturn with nominal doping observed in low-temperature magnetic susceptibility, and further corroborates the presence of intrinsic $Cu^{2+}$ impurities in the doped system as isolated, dimer-free $Cu^{2+}$ spins emerging after the La-doping. Furthermore, at the lowest investigated temperatures, $T \leq 5.5$ K, we observe the emergence of a third, broader component, which can be tentatively associated with the emerging correlations between dimer-free $Cu^{2+}$ spins when their concentration increases.

For a quantitative analysis, the ESR spectra are fitted using up to three Lorentzian components, depending on the temperature range (**Figure 8a**).

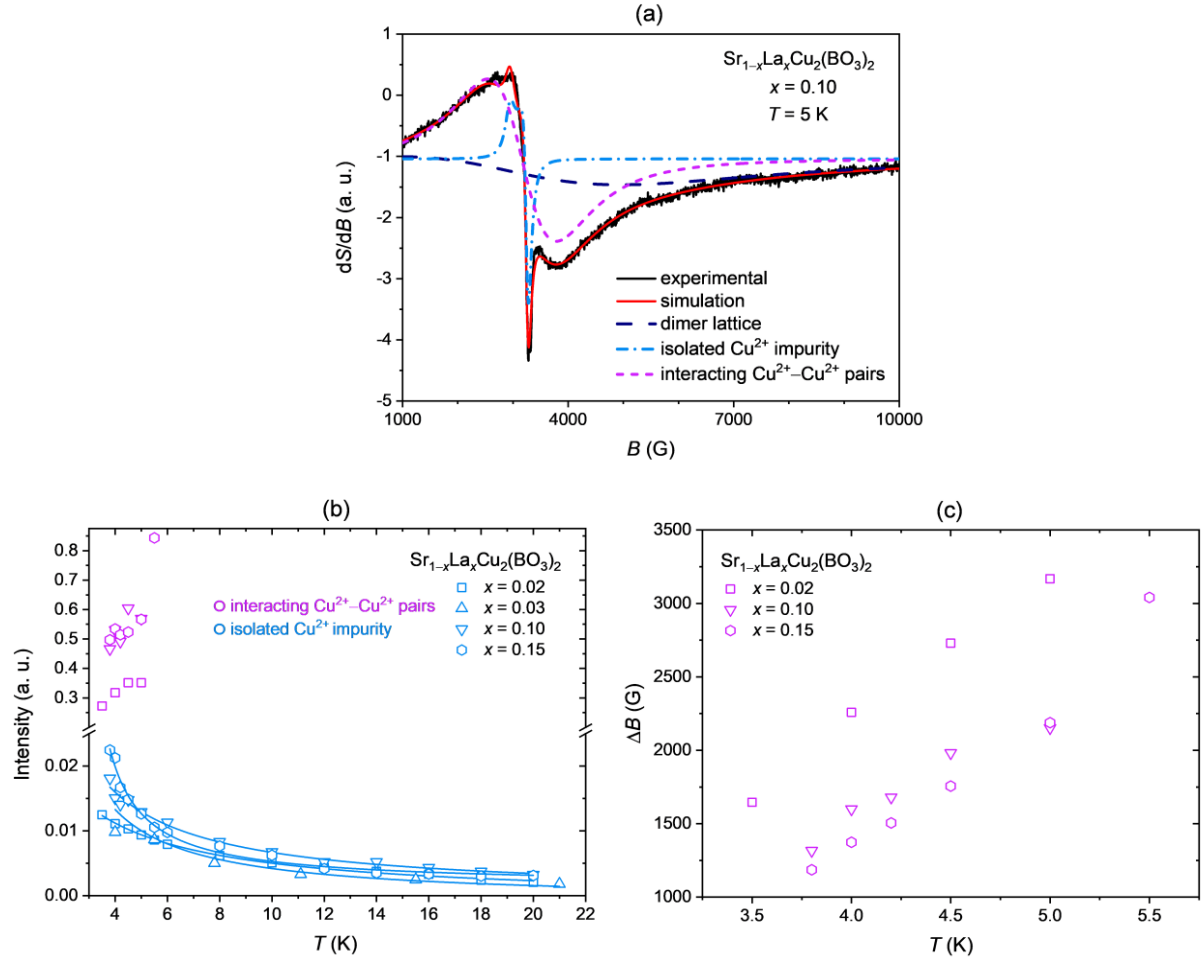

**Figure 8** (**a**) The X-band ESR spectrum of several $Sr_{1-x}La_xCu_2(BO_3)_2$ single crystals with $x = 0.10$ measured at 5 K. The corresponding simulation (red line) is composed of three Lorentzian components: (i) the broad dimer lattice signal (dashed dark blue lines), (ii) an anisotropic component attributed to the isolated $Cu^{2+}$ impurity component (short dash-dot light-blue lines), and (iii) interacting $Cu^{2+}$–$Cu^{2+}$ pairs that emerge between dimer-free $Cu^{2+}$ spins (short dashed violet lines)). (**b**) Temperature dependence of the signal intensities of the isolated $Cu^{2+}$ impurity component and the interacting $Cu^{2+}$–$Cu^{2+}$ pair component for selected nominal doping levels. (**c**) Temperature dependence of the linewidth $\Delta B$ of the interacting $Cu^{2+}$–$Cu^{2+}$ pair component (iii) for $x = 0.02, 0.10$, and $0.15$, showing a monotonic decrease with the decreasing temperature for all doping levels, consistent with the presence of antiferromagnetic exchange interactions between dimer-free $Cu^{2+}$ spins.

At high temperatures, only the broad line attributed to the $Cu^{2+}$ dimer lattice is required. Below 50 K, an additional anisotropic component with the axial $g$-factor anisotropy is included to account for the isolated $Cu^{2+}$ intrinsic impurities. Here, $g_{\parallel}$ and $g_{\perp}$ represent the two $g$-factor eigen-values parallel and perpendicular to the crystallographic $c$ axis, respectively. A corresponding uniaxial anisotropy is also applied to the linewidth, defined by $\Delta B_{\parallel}$ and $\Delta B_{\perp}$ for the field orientations along and perpendicular to the $c$ axis, respectively. Below 5.5 K, a third broad isotropic Lorentzian line has to be introduced, which is tentatively assigned to the onset of emerging correlations between unpaired $Cu^{2+}$ spins. The intensities of all three components, resulting from the fits, are normalized to the room temperature value for all samples.

At room temperature, the main dimer-lattice component is centered around $g \approx 2.1$. The linewidth, $\Delta B$, remains similar across all nominal doping concentrations, ranging from 1350 to 1550 G (**Supplementary Information, Figure S7a**). This matches the corresponding linewidth and $g$-factor values reported in the literature for undoped $SrCu_2(BO_3)_2$ powder.[44] Upon cooling, the linewidth of this component increases monotonically, saturating near 10 K at values between 4400 and 5100 G. This broadening reflects the gradual development of short-range antiferromagnetic correlations in the dimer lattice, and becomes less dominant at low temperatures as the isolated $Cu^{2+}$ impurity component begins to dominate the ESR response. The intensity of this component increases with decreasing temperature (**Figure 8b**), following a $1/T$ dependence across all nominal doping concentrations. The data are fitted using the Curie–Weiss law, yielding a Curie–Weiss temperature $\theta'$ between –2 and +2.5 K, effectively close to zero, confirming that the corresponding $Cu^{2+}$ moments are only weakly interacting and supports their identification as isolated, dimer-free spins. Compared to the isolated $Cu^{2+}$ impurity component, the intensities of the broader component assigned to liberated but interacting $Cu^{2+}$–$Cu^{2+}$ pairs, introduced in the low-temperature regime (3.5–5.5 K), are significantly higher across all doping concentrations (**Figure 8b**). Specifically, the interacting $Cu^{2+}$–$Cu^{2+}$ pair component intensity exceeds the isolated $Cu^{2+}$ impurity component intensity by factors ranging from approximately 10 to 40, highlighting the emergence of magnetic correlations between dimer-free $Cu^{2+}$ spins. These correlations become increasingly more prominent with higher nominal doping. For instance, samples with $x = 0.10$ and 0.15 exhibit stronger intensities than $x = 0.02$, suggesting that higher nominal doping results in more pronounced $Cu^{2+}$–$Cu^{2+}$ interactions. Moreover, at higher nominal doping, these correlated states persist to higher temperatures, as evidenced by the interacting $Cu^{2+}$–$Cu^{2+}$ pair component up to 5.5 K for $x = 0.15$ sample, compared to only up to 5 K for $x = 0.02$ and 0.10. This trend is further corroborated by the temperature dependence of the linewidth, $\Delta B$, of the interacting $Cu^{2+}$–$Cu^{2+}$ pair component.

Overall, $\Delta B$ shows a systematic decrease with increasing doping concentration, as illustrated at 4 K (**Figure 8c**)—where the linewidth reaches its highest value of 2260 G for $x = 0.02$, decreases to 1600 G for $x = 0.10$, and attains its lowest value of 1370 G for $x = 0.15$. Additionally, for each doping concentration individually, the linewidth exhibits a consistent monotonic decrease with the decreasing temperature, indicative of antiferromagnetic nature of the exchange interactions between dimer-free $Cu^{2+}$ spins. The $g$-factor of the interacting $Cu^{2+}$–$Cu^{2+}$ pair component remains constant at approximately 2.12 for all doping levels across the whole temperature range (**Supplementary Information, Figure S7b**).

Due to the large number of fitting parameters and their strong correlations, the linewidth and *g*-factor components of the isolated $Cu^{2+}$ impurity component were partially constrained during analysis. Specifically, $\Delta B_\parallel$ was fixed across all datasets to ensure stable fitting. Throughout most of the temperature range, $\Delta B_\perp$ remained relatively constant, falling within the interval 150–200 G (**Supplementary Information, Figure S7c**). However, below 5.5 K—coinciding with the emergence of component (iii)—a monotonic increase in $\Delta B_\perp$ was observed, providing an additional indication of developing correlations between dimer-free $Cu^{2+}$ spin pairs. Similarly, $g_\parallel$ was fixed for all doping concentrations except for $x = 0.02$, where a gradual decrease was detected below 6 K. The perpendicular component, $g_\perp$, remained effectively constant around 2.065 across the full temperature range for all samples (**Supplementary Information, Figure S7d**).

In contrast to the general trends observed across nominal $x = 0.02$–0.15, the spectra for nominal $x = 0.05$ deviate substantially for both the isolated $Cu^{2+}$ impurity component and the interacting $Cu^{2+}$–$Cu^{2+}$ pair component (**Supplementary Information, Figure S8a**). Contrary to the rest of the samples, this sample exhibits a clear and distinct *g*-factor anisotropy of the isolated $Cu^{2+}$ impurity component, which allows a clear and unambiguous determination of both $\Delta B_\parallel$ and $\Delta B_\perp$. These linewidth components exhibit a substantial increase with decreasing temperature (**Supplementary Information, Figure S8b**), exceeding the modest changes in the linewidth observed for the rest of the samples, which occur only below 5.5 K. The intensity of the isolated dimer-free $Cu^{2+}$ component behaves similarly to the other samples (**Supplementary Information, Figure S8c**), following a 1/*T* dependence and yielding a Curie–Weiss temperature, θ′, close to 0. In contrast, the interacting $Cu^{2+}$–$Cu^{2+}$ pair component for $x = 0.05$ shows further anomalies as its $\Delta B$, remains constant at a significantly lower value of approximately 750 G (**Supplementary Information, Figure S8d**), deviating from the decreasing trend observed in the other samples.

In summary, X-band ESR measurements across all nominal La-doping concentrations reveal multiple components in the spectra: a broad signal from the spin dimer lattice, an anisotropic contribution, which denotes the presence of isolated $Cu^{2+}$ impurities, and an additional low-temperature component arising from dimer-free $Cu^{2+}$–$Cu^{2+}$ interactions. The evolution of the latter component below 5.5 K suggests antiferromagnetic correlations between interacting dimer-free $Cu^{2+}$ spin pairs, which enhance with increasing doping.

### 3.1.5. High magnetic field measurements

While magnetic susceptibility provides an indirect estimate of the spin gap through fitting low-temperature data, high-field ESR allows for its direct determination. The first reported spin gap for $SrCu_2(BO_3)_2$ was Δ = 34 K in 1999.[19] In the present study, to obtain a more precise measurement of the spin gap for La-doped SrCu$_2$(BO$_3$)$_2$, a group of stacked single crystals with nominal $x$ = 0.03 was analyzed using high-field ESR at multiple frequencies up to 500 GHz and magnetic fields up to 16 T (**Figure 9a**). The main signals previously observed within this frequency–field range for the undoped system are two low-lying singlet-triplet excitations, O$_1$ and O$_2$, as well as paramagnetic signals attributed to thermally excited triplets.[48]

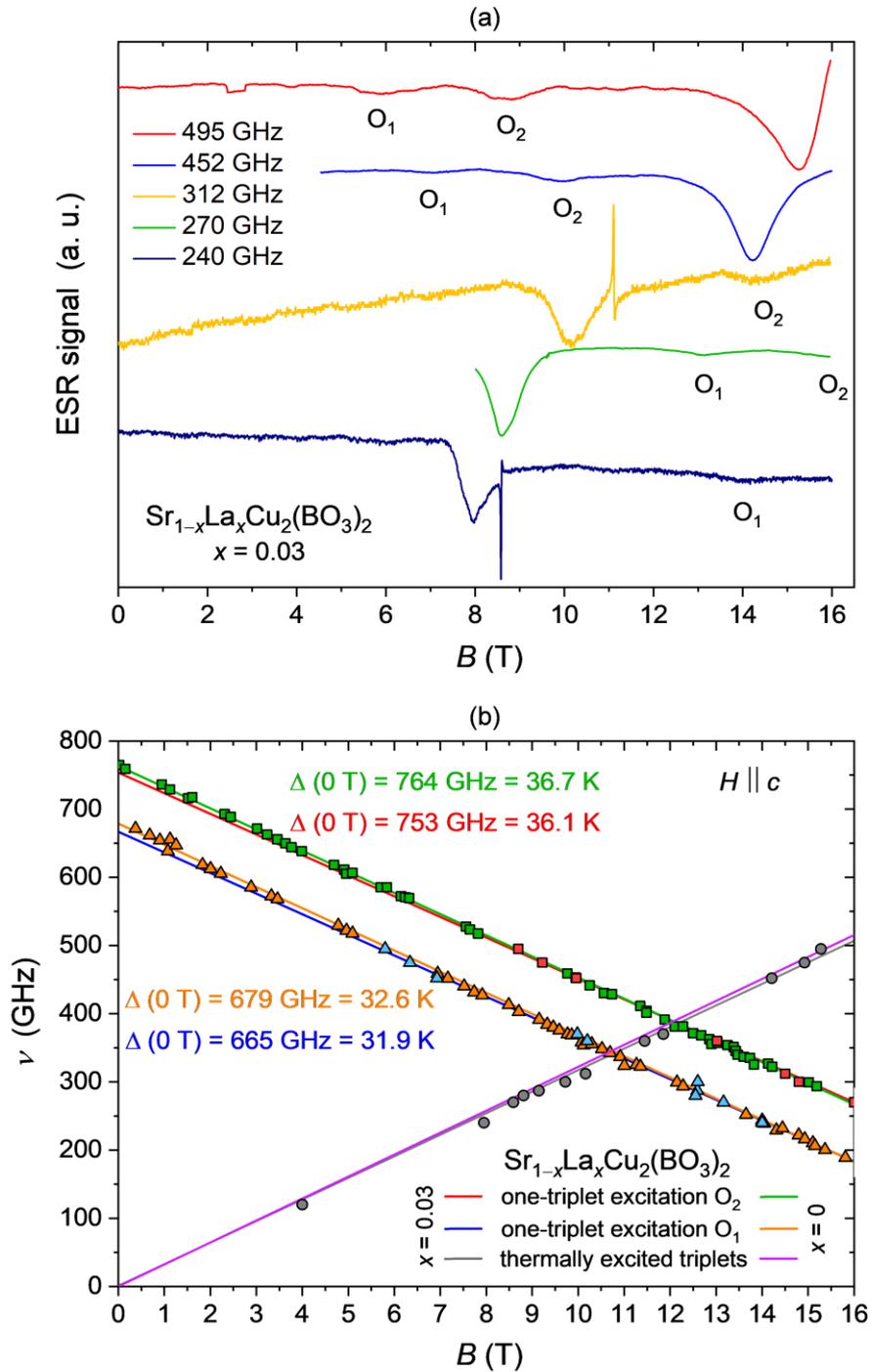

**Figure 9** (a) Low-temperature high-field ESR spectra of stacked Sr$_{1-x}$La$_x$Cu$_2$(BO$_3$)$_2$ single crystals with x = 0.03 at selected frequencies. (b) The corresponding frequency-field diagram ($x$ = 0.03) at $T$ = 1.7–2.9 K for $H \parallel c$, comparing thermally excited triplets and two sets of one-triplet excitations, O$_1$ and O$_2$, with literature data for the undoped system ($x$ = 0), showing a respective 0.6 and 0.7 K reduction of the spin gap induced by doping.

The frequency–field diagram of the undoped system in the orientation $H \parallel c$, reveals zero-field energy gaps of 764(2) GHz and 679(2) GHz for $O_1$ and $O_2$, respectively, corresponding to 36.7 K and 32.6 K.[48] Comparing these values to those obtained experimentally from the frequency-field diagram of the $x$ = 0.03 doped crystals (**Figure 9b**), the spin gaps are only found to be slightly reduced to 753 GHz (36.1 K) for $O_1$ and 665 GHz (31.9 K) for $O_2$. These represent decreases of only 0.6 K and 0.7 K compared to the undoped system, indicating that the effect of nominal $x$ = 0.03 La-doping has a minimal effect on the spin gap of $SrCu_2(BO_3)_2$. Since the $O_1$ and $O_2$ excitations reflect the intrinsic spin dynamics of the dimerized $Cu^{2+}$ lattice and do not account for the presence of intrinsic dimer-free $Cu^{2+}$ impurities or their local magnetic interactions, the observed minimal reduction in the spin gap indicates that the collective spin excitations within the main dimer lattice remain largely unaffected by nominal $x$ = 0.03 doping. Consequently, the notable decrease in the spin gap of 4.7 K, determined from magnetic susceptibility, is primarily influenced by localized magnetic moments associated with intrinsic impurity sites. Capturing additional excitations caused by intrinsic impurities and their interactions through high-field ESR will require measurements at higher frequencies and/or stronger magnetic fields, as well as examination of the remaining samples covering the entire nominal doping range.

## 4. CONCLUSIONS

This study introduces a facile flux method to grow La-doped $SrCu_2(BO_3)_2$ single crystals with lateral sizes up to 3 mm. A large range of La nominal doping concentrations, $x$ = 0.02, 0.03, 0.04, 0.05, 0.10, and 0.15 in $Sr_{1-x}La_xCu_2(BO_3)_2$, is systematically explored. Doping-induced structural changes are subtle but systematic, causing a slight decrease in unit cell volume with increasing La-doping. The successful incorporation of La into the parent $SrCu_2(BO_3)_2$ structure is corroborated by EDS-SEM analysis. A semi-quantitative evaluation of effective doping concentrations reveals that La incorporation reaches approximately 50% of the nominal values for doping levels up to $x$ = 0.05, where the effective doping seems to saturate. This further highlights the inherent difficulty of doping $SrCu_2(BO_3)_2$. In addition to the structural modifications, La-doping induces systematic effects on the magnetism of $SrCu_2(BO_3)_2$. Compared to the undoped system, a progressive closing of the effective spin gap, decreasing from 28.2 K to 20.3 K, is observed for nominal La-doping of $x$ = 0.15. Additionally, a systematic increase of dimer-free $Cu^{2+}$ ($S$ = ½) spin fraction is observed with nominal doping, which also shows antiferromagnetic correlations emerging between dimer-free $Cu^{2+}$ spin pairs below 5.5 K. High-field ESR measurements further indicate that at low doping levels, the spin-gap reduction observed in susceptibility primarily arises from these intrinsic impurity spins, while the collective spin dynamics of the $Cu^{2+}$ dimer lattice remain largely intact. Despite these systematic changes in structure and magnetism, no superconductivity is observed across the entire range of La doping studied.

The present findings underscore the potential of flux growth for La-doping in SrCu$_2$(BO$_3$)$_2$, providing a more efficient and accessible alternative to optical floating zone growth, with the possibility of extending this method to other dopants.

## Supporting Information

The Supporting Information contains: the Rietveld refinement results, average semi-quantitative atomic % values of constituent elements in Sr$_{1-x}$La$_x$Cu$_2$(BO$_3$)$_2$ single crystals, determined by EDS-SEM, compared to nominal values, results of low-temperature fitting results of magnetic susceptibility, and low-temperature inverse susceptibility fits in the interval 2–3.4 K, high- and low-temperature X-band ESR spectra and the development of multiple parameters for three components, derived from the simulations, for selected nominal doping concentrations.

## ACKNOWLEDGEMENTS

The work was supported by Slovenian Research Agency (P2-0105, N1-0397, P1-0125), Young Researcher's Program; Marie Curie Individual Fellowship (Grant No. 101031415), European Union's Horizon 2020 Research and Innovation Program. Part of the research described in this paper was performed at the Canadian Light Source, a national research facility of the University of Saskatchewan, which is supported by the Canada Foundation for Innovation (CFI), the Natural Sciences and Engineering Research Council (NSERC), the Canadian Institutes of Health Research (CIHR), the Government of Saskatchewan, and the University of Saskatchewan.

We also acknowledge support of the Deutsche Forschungsgemeinschaft as well as Dresden High Magnetic Field Laboratory at Helmholtz-Zentrum Dresden-Rossendorf, member of the European Magnetic Field Laboratory, and the Würzburg-Dresden Cluster of Excellence on Complexity and Topology in Quantum Matter – ct.qmat (EXC 2147, project No. 390858490).

# SUPPLEMENTARY INFORMATION FILE

## Influence of La-doping on the magnetic properties of the two-dimensional spin-gapped system $SrCu_2(BO_3)_2$

Lia Šibav[1,2], Tilen Knaflič[1], Graham King[3], Zvonko Jagličić[4,5], Maja Koblar[1], Kirill Povarov[6], Sergei Zvyagin[6], Denis Arčon[1,7], and Mirela Dragomir[1,2*]

[1] Jožef Stefan Institute, Jamova cesta 39, 1000 Ljubljana, Slovenia
[2] Jožef Stefan International Postgraduate School, Jamova cesta 39, 1000 Ljubljana, Slovenia
[3] Canadian Light Source, 44 Innovation Blvd, Saskatoon, SK S7N 2V3, Canada
[4] Institute of Mathematics, Physics and Mechanics, Jadranska ulica 19, 1000 Ljubljana, Slovenia
[5] Faculty of Civil and Geodetic Engineering, University of Ljubljana, Jamova cesta 2, 1000 Ljubljana, Slovenia
[6] Dresden High Magnetic Field Laboratory and Würzburg-Dresden Cluster of Excellence ct.qmat, Helmholtz-Zentrum Dresden-Rossendorf, 01328 Dresden, Germany
[7] Faculty of Mathematics and Physics, University of Ljubljana, Jadranska ulica 19, 1000 Ljubljana, Slovenia

*Corresponding author: mirela.dragomir@ijs.si

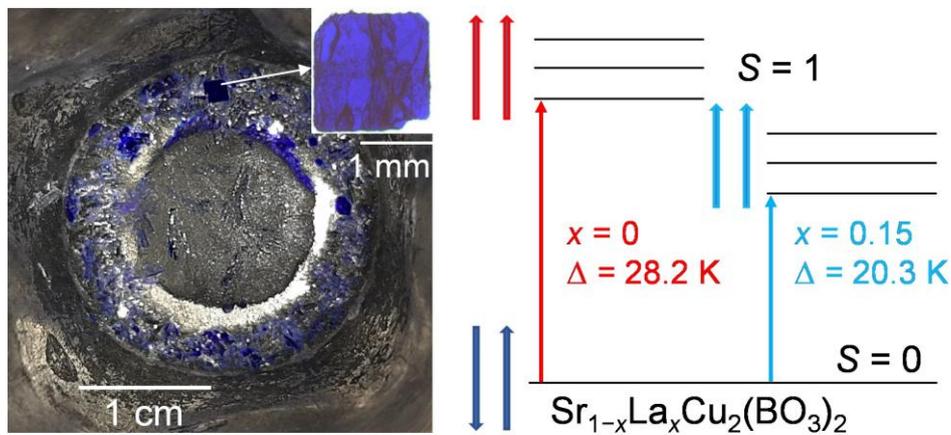

Table of Contents

**Table S1** Room-temperature structural parameters extracted from Rietveld refinement analyses of synchrotron PXRD data collected on crushed $Sr_{1-x}La_xCu_2(BO_3)_2$ single crystals of all studied nominal doping concentrations, $x$ = 0.02–0.15, compared with polycrystalline undoped $SrCu_2(BO_3)_2$ according to ref. [1]. The refinements were performed in the $\bar{I}42m$ space group.

| Nominal $x$ | $a, b$ (Å) | $c$ (Å) | $V$ (Å³) |
|---|---|---|---|
| 0 | 8.9932(1) | 6.6509(1) | 537.91(2) |
| 0.02 | 8.9924(1) | 6.6492(9) | 537.68(1) |
| 0.03 | 8.9885(1) | 6.6468(1) | 537.01(2) |
| 0.04 | 8.9908(4) | 6.6488(3) | 537.46(6) |
| 0.05 | 8.9915 (1) | 6.6495(1) | 537.59(2) |
| 0.10 | 8.9878(1) | 6.6459(1) | 536.85(2) |
| 0.15 | 8.9867(1) | 6.6454(1) | 536.69(1) |

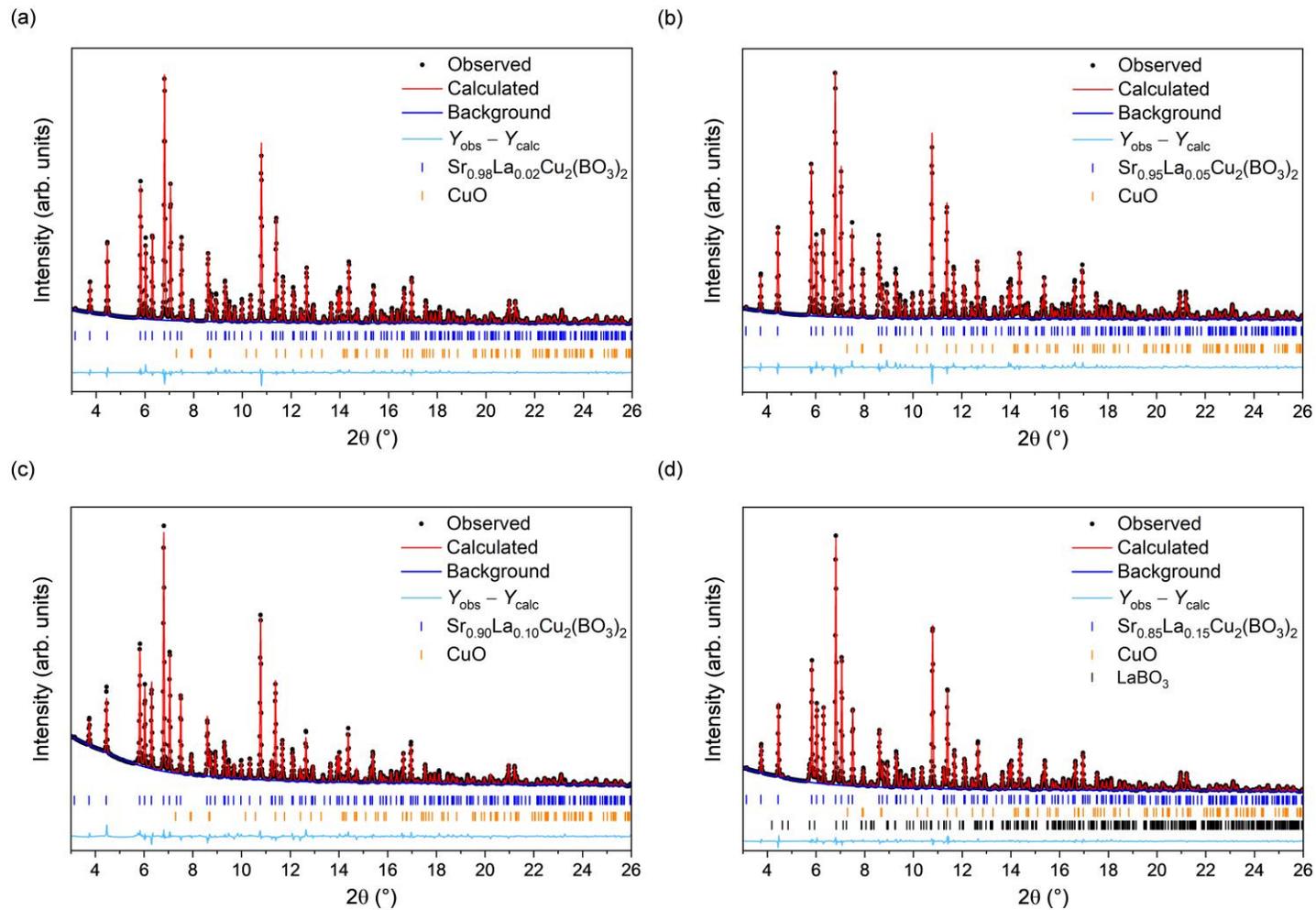

**Figure S1** Rietveld refinement profiles of synchrotron XRD data (λ = 0.3502 Å) collected on crushed $Sr_{1-x}La_xCu_2(BO_3)_2$ single crystals with selected nominal doping concentrations (a) $x$ = 0.02, (b) 0.05, (c) 0.10, and (d) 0.15, collected at room temperature. The refinements were performed using the $I\bar{4}2m$ space group.

Table S2 Detailed results of Rietveld refinement analyses of synchrotron PXRD data collected on crushed $Sr_{1-x}La_xCu_2(BO_3)_2$ single crystals of selected nominal doping concentrations $x$ = 0.02, 0.05, 0.10 and 0.15. The data were refined using the $I\bar{4}2m$ space group.

| | $Sr_{0.98}La_{0.02}Cu_2(BO_3)_2$ | $Sr_{0.95}La_{0.05}Cu_2(BO_3)_2$ | $Sr_{0.90}La_{0.10}Cu_2(BO_3)_2$ | $Sr_{0.85}La_{0.15}Cu_2(BO_3)_2$ |
|---|---|---|---|---|
| $a, b$ (Å) | 8.9924(1) | 8.9915 (1) | 8.9878(1) | 8.9867(1) |
| $c$ (Å) | 6.6492(9) | 6.6495(1) | 6.6459(1) | 6.6454(1) |
| $V$ (Å³) | 537.68(1) | 537.59(2) | 536.85(2) | 536.69(1) |
| $R_{wp}$ (%) | 3.72 | 4.90 | 3.66 | 2.59 |
| Atom | x/y/z<br>Occ.<br>$U_{11}/U_{22}/U_{33}$<br>$U_{12}/U_{13}/U_{23}$ | x/y/z<br>Occ.<br>$U_{11}/U_{22}/U_{33}$<br>$U_{12}/U_{13}/U_{23}$ | x/y/z<br>Occ.<br>$U_{11}/U_{22}/U_{33}$<br>$U_{12}/U_{13}/U_{23}$ | x/y/z<br>Occ.<br>$U_{11}/U_{22}/U_{33}$<br>$U_{12}/U_{13}/U_{23}$ |
| Sr | 0/0.5/0<br>0.99[a]<br>0.0085(9)/0.0026(8)/0.0097(6)<br>0/0/0 | 0/0.5/0<br>0.97[a]<br>0.0077(10)/0.0012(10)/0.0051(7)<br>0/0/0 | 0/0.5/0<br>0.98[a]<br>0.0028(15)/0.0099(17)/0.0071(12)<br>0/0/0 | 0/0.5/0<br>0.97[a]<br>0.0112(8)/0.0071(8)/0.0126(6)<br>0/0/0 |
| La | 0/0.5/0<br>0.01[a]<br>0.0085(9)/0.0026(8)/0.0097(6)<br>0/0/0 | 0/0.5/0<br>0.03[a]<br>0.0077(10)/ 0.0012(10)/0.0051(7)<br>0/0/0 | 0/0.5/0<br>0.02[a]<br>0.0028(15)/0.0099(17)/0.0071(12)<br>0/0/0 | 0/0.5/0<br>0.03[a]<br>0.0112(8)/0.0071(8)/0.0126(6)<br>0/0/0 |
| Cu | 0.11435(6)/0.11435(6)/ 0.27954(15)<br>0.9733(23)<br>0.0015/ 0.0015/0.0170<br>0.0005/0.0014/0.0014 | 0.11463(8)/0.11463(8)/0.28055(20)<br>0.990(4)<br>0.0050/0.0050/0.0177<br>0.0007/0.0031/0.0031 | 0.11465(11)/0.11465(11)/ 0.28038(28)<br>1.024(5)<br>0.0069/0.0069/0.0215<br>0.0050/0.0011/0.0011 | 0.11441(6)/0.11441(6)/0.27939(15)<br>0.9793(26)<br>0.0044/0.0044/0.0245<br>0.0020/0.0020/0.0020 |
| B | 0.2939(5)/0.2939(5)/0.2470(21)<br>1.018(12)<br>0.0061/0.0061/0.0152<br>0.0500/0.0500/0.0500 | 0.2925(7)/0.2925(7)/0.2353(24)<br>1.00[b]<br>0.0050/0.0050/0.0102<br>0.0020/0.0050/0.0050 | 0.2559(9)/0.3169(8)/0.2632(23)<br>1.00[b]<br>0.0050/0.0225/0.0109<br>0.0050/0.0050/0.0050 | 0.2922(5)/0.2922(5)/0.2479(19)<br>0.977(11)<br>0.0100/0.0100/0.0050<br>0.0095/0.0013/0.0013 |
| O1 | 0.4025(3)/0.4025(3)/0.2030(9)<br>0.982(8)<br>0.0113/0.0113/0.0899<br>0.0500/ 0.0244/ 0.0244 | 0.4010(4)/0.4010(4)/0.1994(12)<br>0.962(11)<br>0.0065/0.0065/0.0805<br>0.0050/0.0140/0.0140 | 0.4025(6)/0.4025(6)/0.2109(17)<br>1.00[b]<br>0.0106/0.0106/0.0661<br>0.0050/0.0053/0.0053 | 0.40099(30)/0.40099(30)/0.2090(8)<br>1.008(8)<br>0.0082/0.0082/0.0472<br>0.0020/0.0010/0.0010 |
| O2 | 0.32771(27)/0.14581(29)/ 0.2552(10)<br>1.00[b]<br>0.0065/0.0143/0.0246<br>0.0002/0.0067/0.0050 | 0.3279(3)/0.1459(4)/0.2604(12)<br>1.00[b]<br>0.0050/0.0050/0.0247<br>0.0020/0.0050/0.0058 | 0.3280(5)/0.1381(5)/0.2617(15)<br>0.970(10)<br>0.0050/0.0107/0.0215<br>0.0072/0.0040/0.0050 | 0.32756(27)/0.14657(29)/0.2553(10)<br>1.00[b]<br>0.0166/0.0139/0.0272<br>0.0020/0.0020/0.0020 |

[a] The fractional occupancies of Sr and La were fixed to the effective values, semi-quantitatively determined from SEM-EDS point analyses (Table S3). [b] The occupancies were fixed to the nominal values of 1.

Table S3 The average semi-quantitative atomic % values of Sr, Cu, O and La in Sr$_{1-x}$La$_x$Cu$_2$(BO$_3$)$_2$ single crystals with nominal $x$ = 0–0.15, followed by La/Cu, La/Sr and Cu/Sr ratios, compared to the nominal values. The results were determined from SEM-EDS point analyses (20 points for each doping concentration) performed on the crystals. Due to the low relative atomic mass of boron, it could not be detected by EDS and was therefore omitted from the semi-quantitative analysis.

| Element | Concentration (at%) | | | | | | | | | | | | | |
|---|---|---|---|---|---|---|---|---|---|---|---|---|---|---|
| | $x = 0$ | | $x = 0.02$ | | $x = 0.03$ | | $x = 0.04$ | | $x = 0.05$ | | $x = 0.10$ | | $x = 0.15$ | |
| | Effective | Nominal | Effective | Nominal | Effective | Nominal | Effective | Nominal | Effective | Nominal | Effective | Nominal | Effective | Nominal |
| Sr | 13.9(4) | 11.11 | 13.6(1) | 10.89 | 13.1(7) | 10.78 | 13.1(3) | 10.67 | 13.3(4) | 10.56 | 12.3(2) | 10.00 | 12.3(2) | 9.44 |
| Cu | 28(2) | 22.22 | 28.1(5) | 22.22 | 26(3) | 22.22 | 25(1) | 22.22 | 29(3) | 22.22 | 25.0(8) | 22.22 | 25.6(4) | 22.22 |
| O | 59(2) | 66.67 | 58.2(6) | 66.67 | 61(3) | 66.67 | 62(1) | 66.67 | 58(3) | 66.67 | 62(1) | 66.67 | 61.9(5) | 66.67 |
| La | 0 | 0 | 0.11(1) | 0.22 | 0.16(5) | 0.33 | 0.15(3) | 0.44 | 0.29(6) | 0.56 | 0.23(4) | 1.11 | 0.28(8) | 1.67 |
| Sum, % | 100.9 | 100 | 99.9 | 100 | 100.1 | 100 | 100.1 | 100 | 100.3 | 100 | 99.5 | 100 | 100.1 | 100 |
| La/Cu | 0 | 0 | 0.0040(3) | 0.01 | 0.006(2) | 0.015 | 0.006(1) | 0.02 | 0.010(2) | 0.025 | 0.009(2) | 0.05 | 0.011(3) | 0.075 |
| La/Sr | 0 | 0 | 0.0082(7) | 0.02 | 0.012(3) | 0.03 | 0.011(2) | 0.04 | 0.022(5) | 0.05 | 0.019(3) | 0.11 | 0.023(6) | 0.18 |
| Cu/Sr | 1.99(9) | 2 | 2.06(2) | 2.04 | 2.0(1) | 2.06 | 1.91(4) | 2.08 | 2.2(2) | 2.10 | 2.02(4) | 2.22 | 2.07(3) | 2.35 |
| La mol% | 0 | 0 | 1.0(1) | 2 | 1.4(5) | 3 | 1.3(2) | 4 | 2.6(5) | 5 | 2.4(3) | 10 | 2.5(7) | 15 |

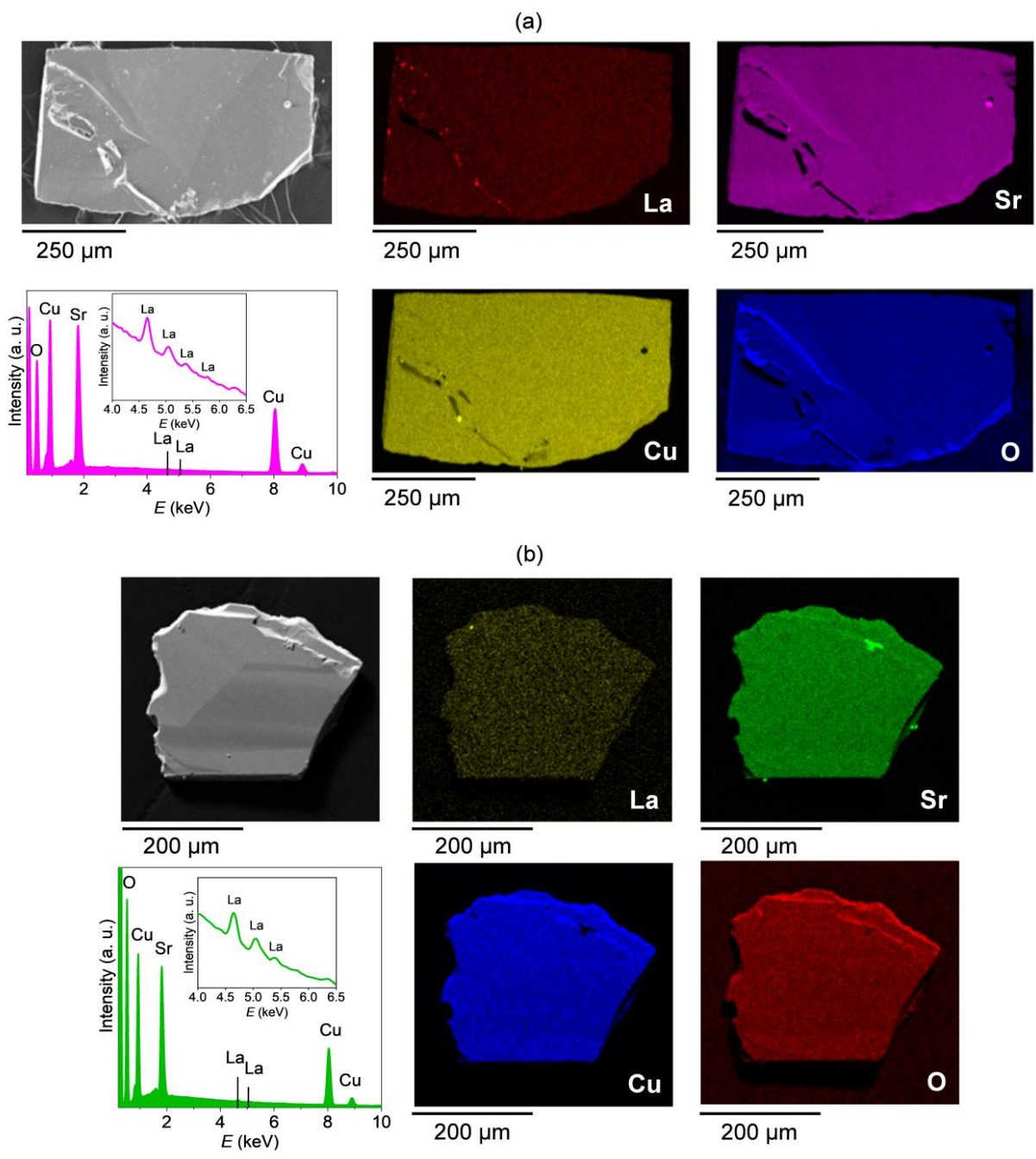

Figure S2 An SEM image and EDS elemental mapping analysis of a Sr$_{1-x}$La$_x$Cu$_2$(BO$_3$)$_2$ single crystal with nominal (a) $x$ = 0.04 and (b) 0.10 reveal a relatively homogeneous distribution of La within the SrCu$_2$(BO$_3$)$_2$ matrix. Few LaBO$_3$ impurity particles are observed on the surface. The corresponding EDS sum spectra clearly show the presence of La emission lines for both a) and b) samples. The insets show a magnified region where the La emission lines are distinctly visible. Few CuO impurity particles are additionally visible in a).

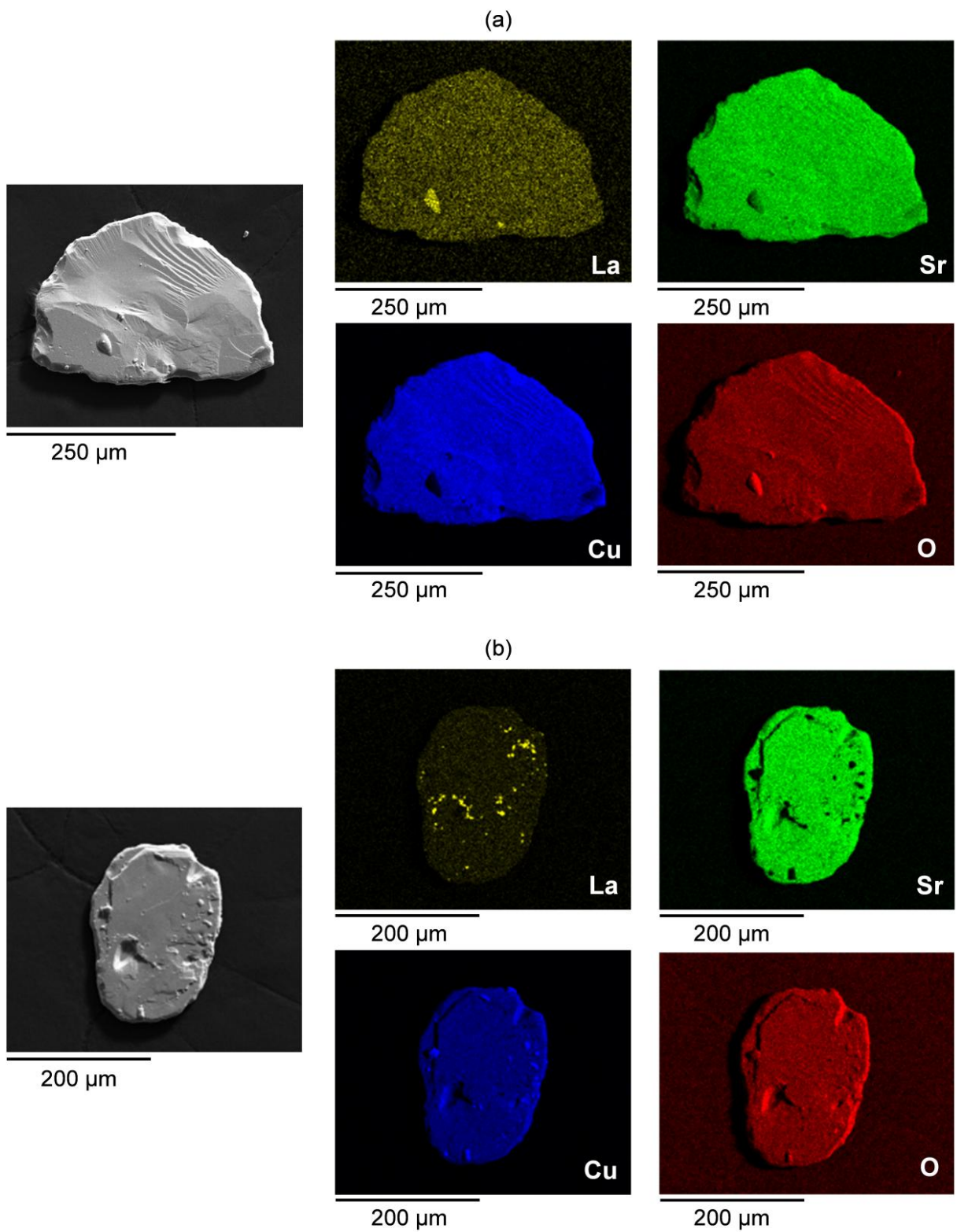

Figure S3 An SEM image (left) and EDS elemental mapping analysis (right) of a $Sr_{1-x}La_xCu_2(BO_3)_2$ single crystal with nominal $x$ = 0.15, showing: (a) the presence of $LaBO_3$ impurity particles on the crystal surface and (b) the presence of both CuO and $LaBO_3$ impurity particles on the crystal surface.

Table S4 Results of the low-temperature magnetic susceptibility data fits (2–6 K range) for Sr$_{1-x}$La$_x$Cu$_2$(BO$_3$)$_2$ single crystals with nominal $x$ = 0–0.15 (Equation 3 in the main text). Five fixed θ' values were used due to its high correlation with C'.

| Nominal $x$ | C' (emu K/mol) | $\chi_0$ (emu/mol) | θ' | Δ (K) | $a$ (emu/mol) |
|---|---|---|---|---|---|
| $x$ = 0 | 3.4(1) · 10$^{-4}$ | 7.1(6) · 10$^{-5}$ | 0 | 28.7(3) | 8.3(4) · 10$^{-2}$ |
| | 4.1(2) · 10$^{-4}$ | 5.7(6) · 10$^{-5}$ | −0.25 | 28.5(3) | 8.1(4) · 10$^{-2}$ |
| | 4.9(2) · 10$^{-4}$ | 4.4(6) · 10$^{-5}$ | −0.5 | 28.3(3) | 7.9(4) · 10$^{-2}$ |
| | 5.7(2) · 10$^{-4}$ | 3.1(7) · 10$^{-5}$ | −0.75 | 28.2(3) | 7.8(3) · 10$^{-2}$ |
| | 6.6(3) · 10$^{-4}$ | 1.8(7) · 10$^{-5}$ | −1 | 28.1(3) | 7.7(3) · 10$^{-2}$ |
| $x$ = 0.02 | 1.95(1) · 10$^{-3}$ | 3.09(6) · 10$^{-4}$ | 0 | 26.7(3) | 6.6(3) · 10$^{-2}$ |
| | 2.37(1) · 10$^{-3}$ | 2.31(5) · 10$^{-4}$ | −0.25 | 25.9(2) | 5.9(2) · 10$^{-2}$ |
| | 2.82(2) · 10$^{-3}$ | 1.52(6) · 10$^{-4}$ | −0.5 | 25.2(2) | 5.4(2) · 10$^{-2}$ |
| | 3.32(2) · 10$^{-3}$ | 7.2(7) · 10$^{-5}$ | −0.75 | 24.7(2) | 5.1(2) · 10$^{-2}$ |
| | 3.86(3) · 10$^{-3}$ | −9(9) · 10$^{-6}$ | −1 | 24.2(2) | 4.8(2) · 10$^{-2}$ |
| $x$ = 0.03 | 2.46(2) · 10$^{-3}$ | 3.83(8) · 10$^{-4}$ | 0 | 25.3(3) | 5.4(2) · 10$^{-2}$ |
| | 2.99(2) · 10$^{-3}$ | 2.82(6) · 10$^{-4}$ | −0.25 | 24.4(2) | 4.8(1) · 10$^{-2}$ |
| | 3.58(2) · 10$^{-3}$ | 1.79(6) · 10$^{-4}$ | −0.5 | 23.6(2) | 4.4(1) · 10$^{-2}$ |
| | 4.18(3) · 10$^{-3}$ | 8.6(8) · 10$^{-4}$ | −0.75 | 23.2(2) | 4.2(1) · 10$^{-2}$ |
| | 4.91(3) · 10$^{-3}$ | −3.2(9) · 10$^{-6}$ | −1 | 22.4(2) | 3.8(1) · 10$^{-2}$ |
| $x$ = 0.04 | 2.25(2) · 10$^{-3}$ | 3.65(9) · 10$^{-4}$ | 0 | 25.4(4) | 5.0(3) · 10$^{-2}$ |
| | 2.74(3) · 10$^{-3}$ | 2.73(9) · 10$^{-4}$ | −0.25 | 24.5(3) | 4.5(2) · 10$^{-2}$ |
| | 3.27(3) · 10$^{-3}$ | 1.8(1) · 10$^{-4}$ | −0.5 | 23.7(3) | 4.1(2) · 10$^{-2}$ |
| | 3.86(4) · 10$^{-3}$ | 8(1) · 10$^{-5}$ | −0.75 | 23.1(3) | 3.8(2) · 10$^{-2}$ |
| | 4.49(5) · 10$^{-3}$ | −1(1) · 10$^{-5}$ | −1 | 22.5(3) | 3.6(2) · 10$^{-2}$ |
| $x$ = 0.05 | 2.39(2) · 10$^{-3}$ | 4.12(6) · 10$^{-4}$ | 0 | 24.8(3) | 4.4(2) · 10$^{-2}$ |
| | 2.90(1) · 10$^{-3}$ | 3.17(5) · 10$^{-4}$ | −0.25 | 23.6(2) | 3.7(1) · 10$^{-2}$ |
| | 3.46(2) · 10$^{-3}$ | 2.16(5) · 10$^{-4}$ | −0.5 | 22.8(2) | 3.37(8) · 10$^{-2}$ |
| | 4.09(2) · 10$^{-3}$ | 1.13(7) · 10$^{-4}$ | −0.75 | 22.1(2) | 3.12(8) · 10$^{-2}$ |
| | 4.68(3) · 10$^{-3}$ | 3.4(8) · 10$^{-5}$ | −1 | 22.0(2) | 3.15(9) · 10$^{-2}$ |
| $x$ = 0.10 | 2.50(2) · 10$^{-3}$ | 5.66(7) · 10$^{-4}$ | 0 | 23.8(3) | 3.0(1) · 10$^{-2}$ |
| | 3.04(1) · 10$^{-3}$ | 4.60(5) · 10$^{-4}$ | −0.25 | 22.6(2) | 2.60(7) · 10$^{-2}$ |
| | 3.64(1) · 10$^{-3}$ | 3.51(4) · 10$^{-4}$ | −0.5 | 21.6(1) | 2.33(4) · 10$^{-2}$ |
| | 4.31(2) · 10$^{-3}$ | 2.39(5) · 10$^{-4}$ | −0.75 | 20.8(1) | 2.14(4) · 10$^{-2}$ |
| | 5.04(2) · 10$^{-3}$ | 1.26(7) · 10$^{-5}$ | −1 | 20.1(2) | 2.00(5) · 10$^{-2}$ |
| $x$ = 0.15 | 2.51(2) · 10$^{-3}$ | 7.3(1) · 10$^{-4}$ | 0 | 22.8(4) | 2.8(2) · 10$^{-2}$ |
| | 3.05(3) · 10$^{-3}$ | 6.3(1) · 10$^{-4}$ | −0.25 | 21.8(4) | 2.5(1) · 10$^{-2}$ |
| | 3.64(4) · 10$^{-3}$ | 5.2(1) · 10$^{-4}$ | −0.5 | 21.0(4) | 2.3(1) · 10$^{-2}$ |
| | 4.30(5) · 10$^{-3}$ | 4.1(2) · 10$^{-4}$ | −0.75 | 20.3(4) | 2.2(1) · 10$^{-2}$ |
| | 5.02(6) · 10$^{-3}$ | 3.0(2) · 10$^{-4}$ | −1 | 19.7(4) | 2.0(1) · 10$^{-2}$ |

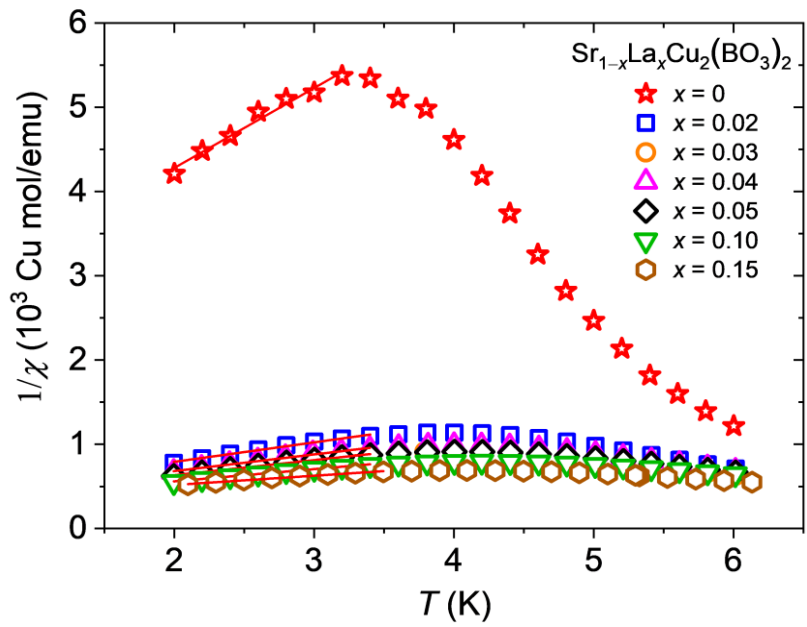

**Figure S4** Inverse magnetic susceptibility data for $Sr_{1-x}La_xCu_2(BO_3)_2$ single crystals with nominal $x$ = 0–0.15, displaying low-temperature Curie–Weiss fits performed in the 2–3.4 K interval to estimate intrinsic $Cu^{2+}$ ($S$ = ½) impurities.

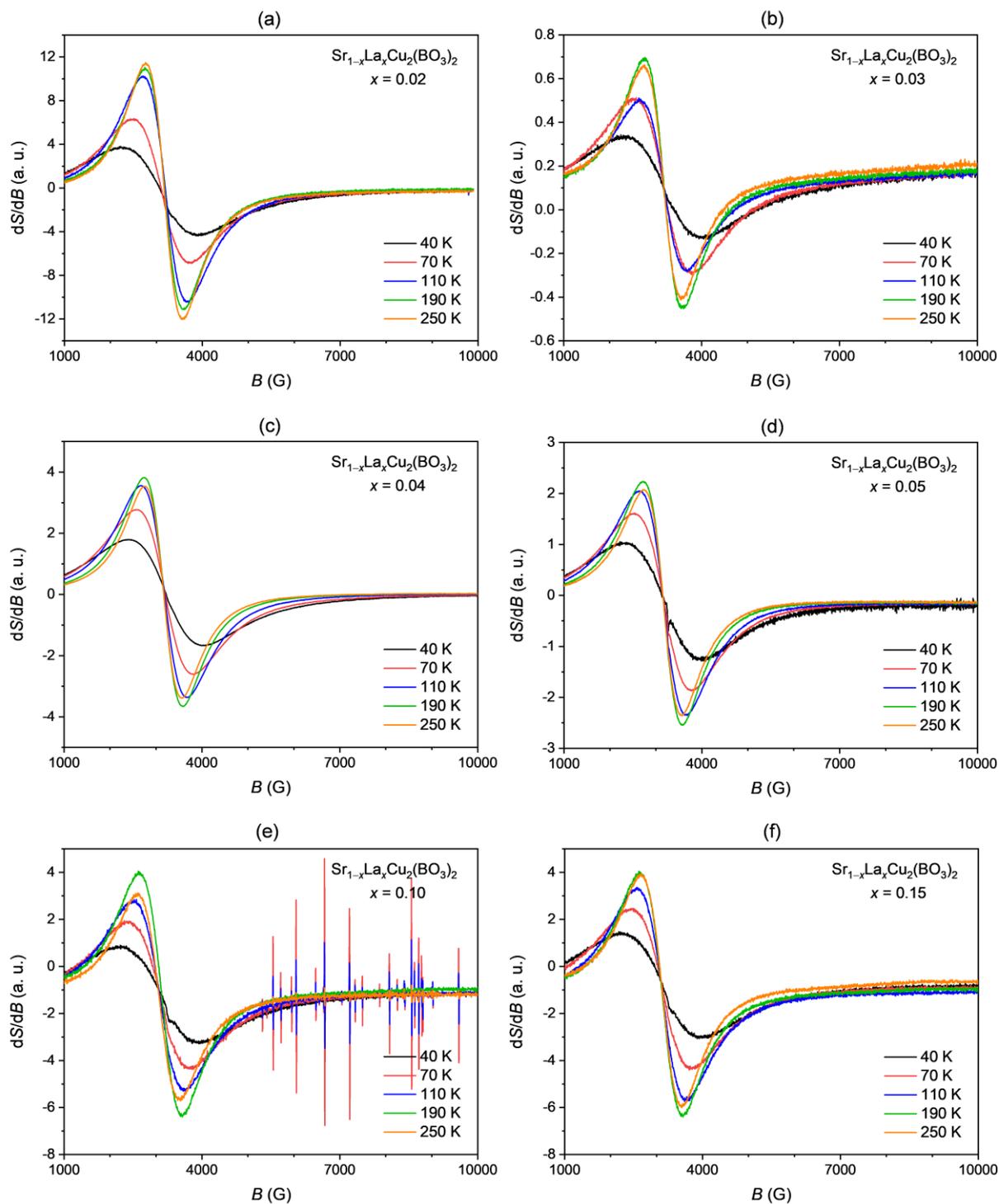

Figure S5 X-band ESR spectra of $Sr_{1-x}La_xCu_2(BO_3)_2$ single crystals with nominal (a) $x$ = 0.02, (b) 0.03, (c) 0.04, (d) 0.05, (e) 0.10 and (f) 0.15, measured as quasi-polycrystalline samples at temperatures 40, 70, 110, 170 and 250 K, which show the dominant signal of the dimer lattice at high temperatures. Additional narrow absorption lines for $x$ = 0.10 sample indicate the presence of an extrinsic paramagnetic impurity, not observed in PXRD or EDS-SEM.

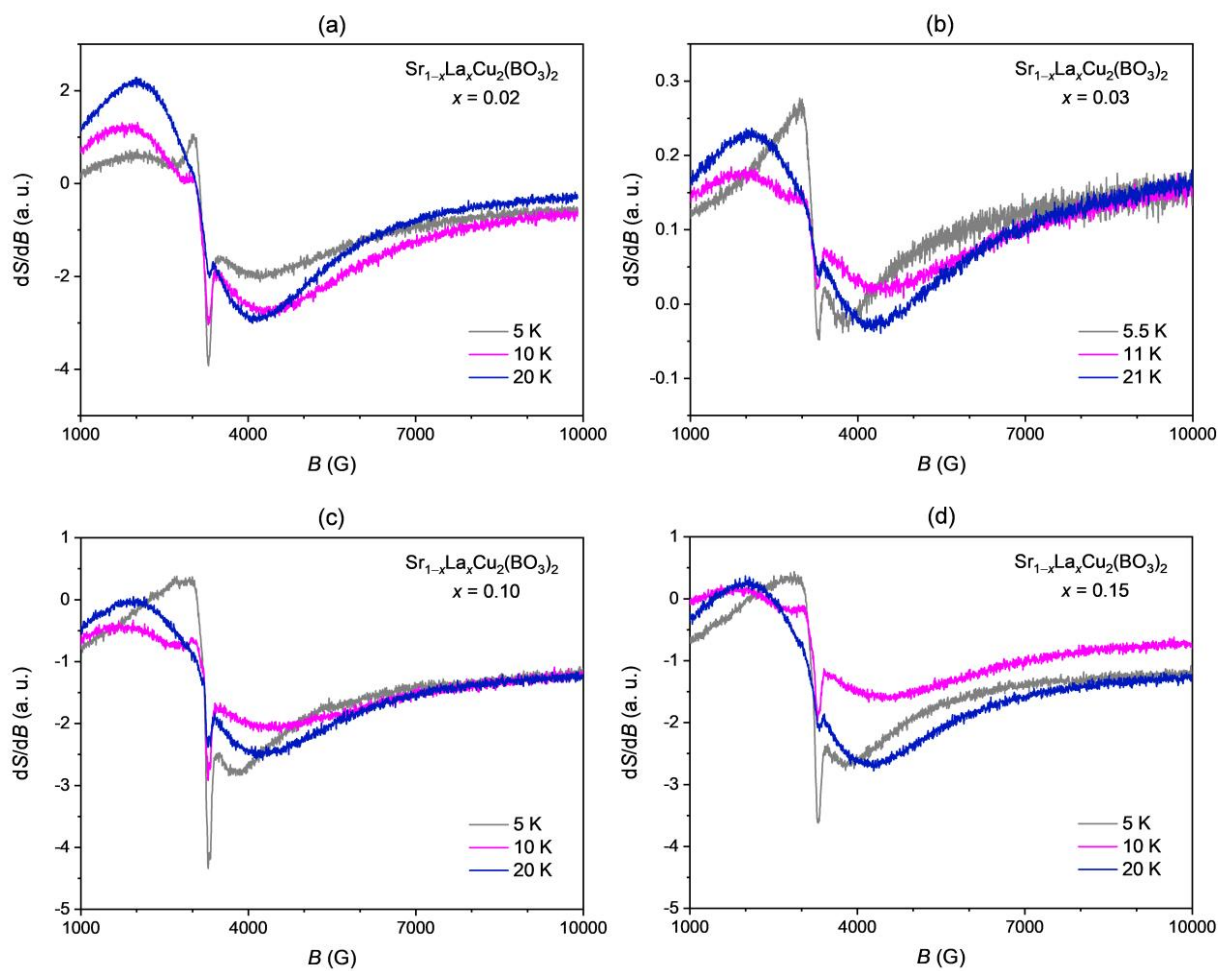

Figure S6 Low-temperature X-band ESR spectra of $Sr_{1-x}La_xCu_2(BO_3)_2$ single crystals with nominal (a) $x$ = 0.02, (b) 0.03, (c) 0.10, and (d) 0.15 at selected temperatures 5, 10 and 20 K, showing the development of the characteristic impurity signal correlated to intrinsic impurities–isolated dimer free $Cu^{2+}$ spins. Comparing the spectra at 5 and 10 K, an additional broadening is observed at 5 K, indicating the presence of an extra component associated with interactions between dimer-free $Cu^{2+}$–$Cu^{2+}$ pairs.

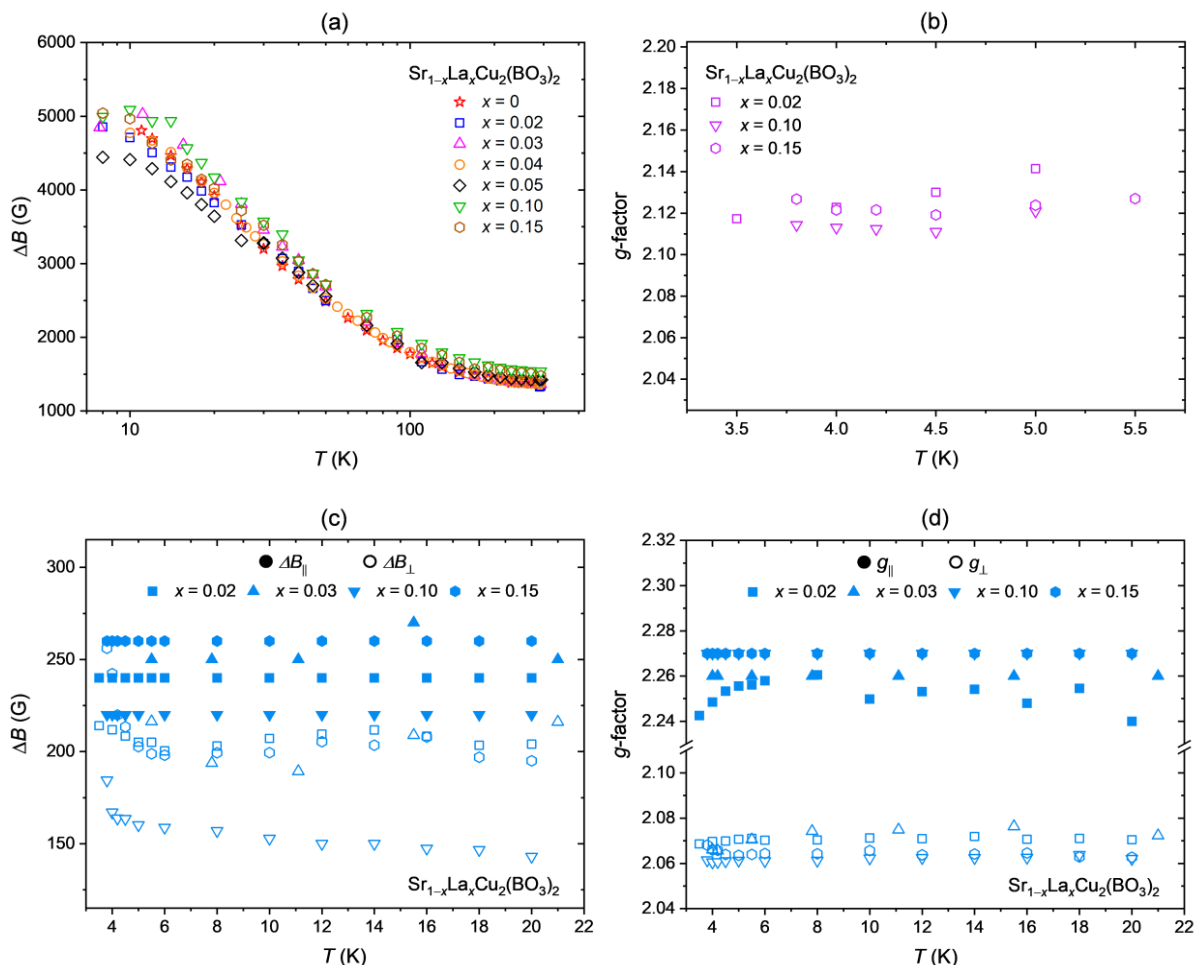

**Figure S7** (**a**) The logarithmic temperature behavior of the X-band ESR linewidth of the main dimer-lattice signal. (**b**) The temperature behavior of $g$-factor for the interacting $Cu^{2+}$–$Cu^{2+}$ pair component for $Sr_{1-x}La_xCu_2(BO_3)_2$ single crystals with nominal $x$ = 0.02, 0.10, and 0.15. (**c**) The temperature behavior of X-band ESR linewidth for the isolated $Cu^{2+}$ impurity component, parallel ($\Delta B_{\parallel}$) and perpendicular ($\Delta B_{\perp}$) to $c$ crystallographic axis for $Sr_{1-x}La_xCu_2(BO_3)_2$ single crystals with nominal $x$ = 0.02, 0.03, 0.10, and 0.15. (**d**) The temperature behavior of $g$-factor for the isolated $Cu^{2+}$ impurity component, parallel ($g_{\parallel}$) and perpendicular ($g_{\perp}$) to $c$ crystallographic axis for $Sr_{1-x}La_xCu_2(BO_3)_2$ single crystals with nominal $x$ = 0.02, 0.03, 0.10, and 0.15.

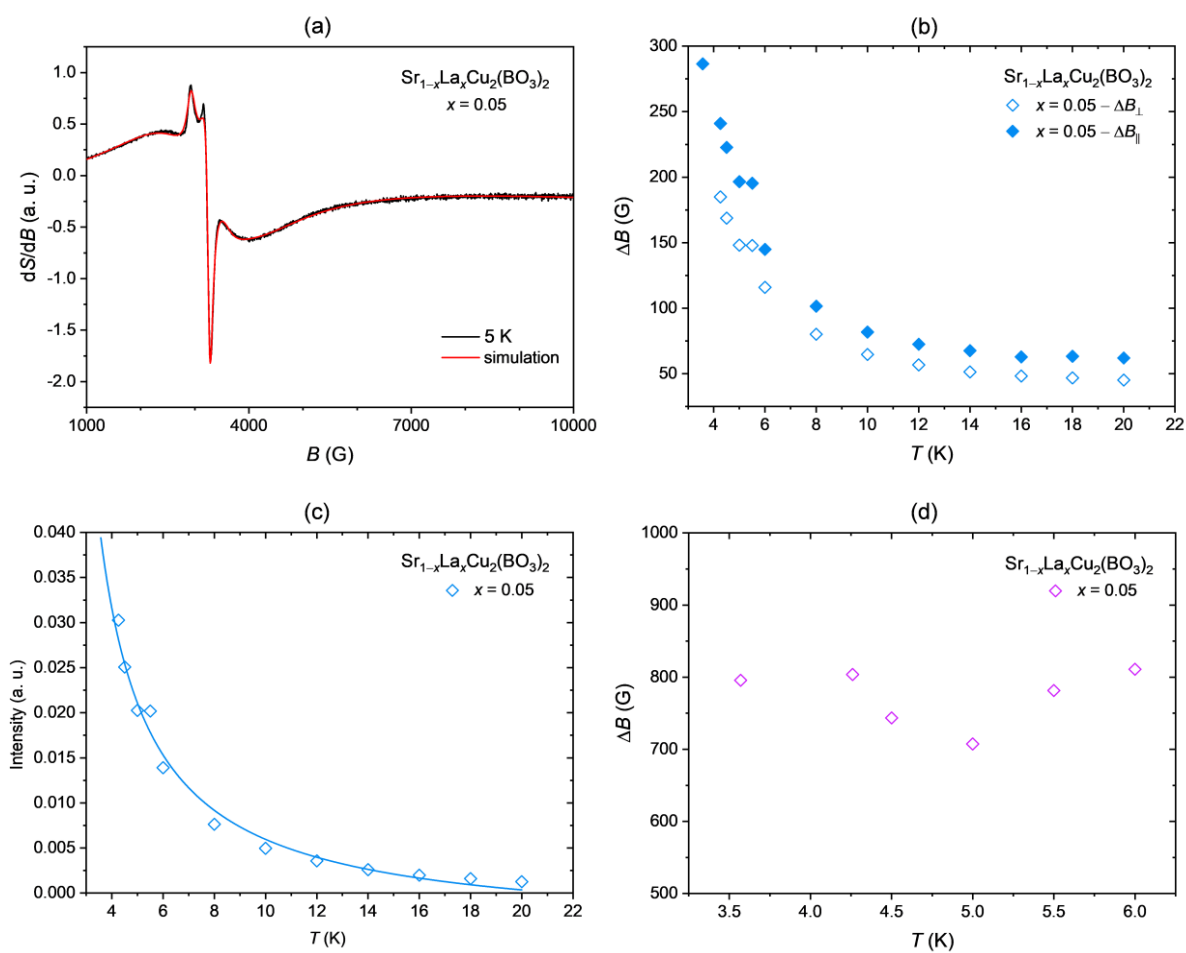

Figure S8 (a) X-band ESR spectrum of $Sr_{1-x}La_xCu_2(BO_3)_2$ single crystals with nominal $x$ = 0.05 at 5 K, showing a clear uniaxial anisotropy opposed to samples with other nominal doping concentrations, and its simulation as a sum of three components. (b) The temperature behavior of X-band ESR linewidth for the isolated $Cu^{2+}$ impurity component, parallel ($\Delta B_{\parallel}$) and perpendicular ($\Delta B_{\perp}$) to $c$ crystallographic axis for nominal $x$ = 0.05, showing a significant increase in both crystallographic directions with the decreasing temperature, contrary to the rest of the samples. (c) Temperature dependence of the signal intensity of the isolated $Cu^{2+}$ impurity component for nominal $x$ = 0.05, which follows a Curie–Weiss law, giving a low Curie-Weiss temperature. This indicates minimal spin correlations, aligning with the rest of the samples. (d) The temperature behavior of X-band ESR linewidth for the interacting $Cu^{2+}$–$Cu^{2+}$ pair component for nominal $x$ = 0.05. Contrary to the rest of the samples, the linewidth remains constant and does not show a decrease with the decreasing temperature.

## REFERENCES

[1] Šibav, L.; Gosar, Ž.; Knaflič, T.; Jagličić, Z.; King, G.; Nojiri, H.; Arčon, D.; Dragomir, M. Higher-Magnesium-Doping Effects on the Singlet Ground State of the Shastry-Sutherland $SrCu_2(BO_3)_2$. *Inorg. Chem.* **2024**, 63, 20335–20346. DOI: 10.1021/acs.inorgchem.4c02398